%% file: main.tex
\documentclass[a4paper,numberwithinsect,USenglish]{lipics-v2021}

\title{On the Parallel Parameterized Complexity\\ of MaxSAT Variants}

\titlerunning{On the Parallel Parameterized Complexity of MaxSAT Variants}

\author{Max Bannach}{Institute for Theoretical Computer Science, Universität zu Lübeck, Germany}{bannach@tcs.uni-luebeck.de}{https://orcid.org/0000-0002-6475-5512}{}
\author{Malte Skambath}{Department of Computer Science, Kiel University, Germany}{malte.skambath@email.uni-kiel.de}{https://orcid.org/0000-0003-2048-3559}{}
\author{Till Tantau}{Institute for Theoretical Computer Science, Universität zu Lübeck, Germany}{tantau@tcs.uni-luebeck.de}{}{}

\authorrunning{M. Bannach,  M. Skambath, and T. Tantau}
\Copyright{M. Bannach, M. Skambath, and T. Tantau} 


\category{} 

\relatedversion{} 

\nolinenumbers 

\hideLIPIcs  

\EventEditors{John Q. Open and Joan R. Access}
\EventNoEds{2}
\EventLongTitle{42nd Conference on Very Important Topics (CVIT 2016)}
\EventShortTitle{CVIT 2016}
\EventAcronym{CVIT}
\EventYear{2016}
\EventDate{December 24--27, 2016}
\EventLocation{Little Whinging, United Kingdom}
\EventLogo{}
\SeriesVolume{42}
\ArticleNo{23}

\theoremstyle{plain}
\newtheorem{problem}[theorem]{Problem}
\newtheorem{fact}[theorem]{Fact}
\newtheorem{reductionrule}{Rule}

\def\reduce{\leq_{\mathrm m}^{\mathrm{para\text-AC^0}}}
\DeclareMathOperator{\vars}{\mathrm{vars}}
\DeclareMathOperator{\clauses}{\mathrm{clauses}}

\usepackage{xfrac}
\usepackage{tikz}
\usetikzlibrary{graphs, arrows.meta}

\definecolor{ba.yellow}{RGB}{252,190,18}
\definecolor{ba.gray}{RGB}{153,153,156}
\definecolor{ba.blue}{RGB}{6,123,164}
\definecolor{ba.red}{RGB}{213,96,98}
\definecolor{ba.orange}{RGB}{233,116,81}
\definecolor{ba.pine}{RGB}{67,154,134}
\definecolor{ba.green}{RGB}{196,247,161}
\definecolor{ba.violet}{RGB}{88, 53, 94}

\definecolor{myred}{rgb}{0.89, 0.26, 0.2}
\definecolor{mygreen}{rgb}{0.31, 0.78, 0.47}

\def\nodetype#1{{\sffamily#1}}
\def\technique#1{\hbox to 0pt{\color{ba.blue}\it Can Be Solved Using~#1\hss}}

\newcommand\classlabel[4][white]{%
  \node[anchor=west, baseline, color=#1] at (0,#2+0.5)          {\small #3};
  \node[anchor=east, baseline, color=#1] at (\linewidth,#2+0.5) {\small #4};
}
\def\coloredclass#1#2#3#4{%
  \fill[color=#2!#3!black!70, rounded corners=6] (0,#1) rectangle (\linewidth,#1+#4);%
}

\usepackage{pifont}
\def\check{\textcolor{ba.green!50!black}{\ding{51}}}
\def\uncheck{\textcolor{ba.red!75!black}{\ding{55}}}

\newcommand\Para{\mathrm{para\text-}}

\newcommand\Class[1]{%
  \mathchoice%
  {\text{\normalfont\small$\mathrm{#1}$}}%
  {\text{\normalfont\small$\mathrm{#1}$}}%
  {\text{\normalfont$\mathrm{#1}$}}%
  {\text{\normalfont$\mathrm{#1}$}}%
}

\newenvironment{parameterizedproblem}%
{%
  \leavevmode\nobreak\par
  \begin{list}%
    {}%
    {%
      \def\labelstyle{\itshape}
      \setlength{\topsep}{0pt}%
      \settowidth{\labelwidth}{\labelstyle Parameter:}%
      \setlength{\leftmargin}{\labelwidth}%
      \addtolength{\leftmargin}{\labelsep}%
      \setlength{\itemsep}{0pt}%
      \setlength{\parsep}{0pt}%
    }%
      \def\instance{\item[\labelstyle Instance:]}%
      \def\parameter{\item[\labelstyle Parameter:]}%
      \def\question{\item[\labelstyle Question:]}%
      \def\promise{\item[\labelstyle Promise:]}%
    }%
    {%
  \end{list}%
}

\newcommand{\Lang}[1]{\text{\normalfont\textsc{#1}}}
\newcommand{\PLang}[2][]{\mathrm{p}_{#1}\Lang{-#2}}

\newcommand{\LP}[1]{\Pi_{\mathrm{#1}}}

\usepackage[nomove]{tcs-automove}
\tcsautomovetheoremenvironment{lemma}{Lemma}
\tcsautomovetheoremenvironment{theorem}{Theorem}
\tcsautomovefurtherenvironment{figure}
\tcsautomovefurtherenvironment{table}

\newcommand{\weightededge}[4][]{%
  \draw[semithick, #1] (#2) to node[midway, circle, inner sep=0.5pt, fill=white] {\tiny\color{ba.violet}$#4$} (#3);%
}
\newcommand{\arc}[4][]{%
  \draw[semithick, ->, >={[round, sep]Stealth}, #1] (#2) to node[near start, circle, inner sep=0.5pt, fill=white] {\tiny\color{ba.violet}$#4$} (#3);%
}
\newcommand{\arcclose}[4][]{%
  \draw[semithick, ->, >={[round, sep]Stealth}, #1] (#2) to node[very near start, circle, inner sep=0.5pt, fill=white] {\tiny\color{ba.violet}$#4$} (#3);%
}

\usepackage{listings}
\lstdefinelanguage{pseudocode}{
  morekeywords={
    algorithm,method,new,and,
    if,then,else,while,do,repeat,until,seq,
    seqdo,return,call,
    for,pardo,foreach,print,output,input,exit,
    break,loop,end,begin,goto,par,global,local,
    read,write,stop,idle,procedure,function,
    throw,catch
  },
  sensitive=true,
  morecomment=[l]{//},
  morestring=[b]",
  morestring=[s]{``}{''},
}
\lstdefinestyle{pseudocode}{
  language=pseudocode,
  basicstyle=\small\rmfamily,
  commentstyle=\upshape\color{black!50},
  keywordstyle=\bfseries\itshape\color{ba.blue},
  identifierstyle=\itshape,
  stringstyle=\rmfamily,
  columns=fullflexible,
  mathescape,
  literate={<-}{{$\gets$\ }}2,
  numbers=left,
  numberstyle=\scriptsize\sffamily,
}

\begin{document}

\maketitle

\begin{abstract}
  In the maximum satisfiability problem (\textsc{max-sat}) we are given a propositional
  formula in conjunctive normal form and have to find an assignment
  that satisfies as many clauses  as possible. We study
  the \emph{parallel} parameterized complexity of various versions of
  \textsc{max-sat} and provide the first \emph{constant-time} algorithms
  parameterized either by the solution size or by the allowed
  excess relative to some guarantee (``above guarantee''
  versions). For the \emph{dual parameterized} version where the 
  parameter is the  number of clauses we are allowed to leave
  unsatisfied, we present 
  the first parallel algorithm for \textsc{max-2sat} (known as \textsc{almost-2sat}).  The difficulty in solving
  \textsc{almost-2sat} in \emph{parallel} comes from the fact that the
  iterative compression method, originally developed to prove that the
  problem is fixed-parameter tractable at all, is inherently
  \emph{sequential.} We observe that a graph flow whose value is a parameter can be
  computed in parallel and use this fact to develop a parallel algorithm
  for the vertex cover problem parameterized above the size of a given
  matching.
  Finally, we study the parallel complexity of \textsc{max-sat}
  parameterized by the vertex cover 
  number, the treedepth, the feedback vertex set number, and the
  treewidth of the input's incidence graph. While \textsc{max-sat} is
  fixed-parameter tractable for all of these parameters, we show that they
  allow different degrees of possible 
  parallelization. For all four we develop dedicated parallel
  algorithms that are \emph{constructive}, meaning that they 
  output an optimal assignment~--~in contrast to results that can be
  obtained by parallel meta-theorems, which often only solve the decision
  version.  
\end{abstract}

{
  \vfill
  \small 
  \tableofcontents

}
\clearpage

\section{Introduction}

Maximum satisfiability problems ask us to find solutions for 
constraint systems that satisfy as many constraints as possible. The
perhaps best-studied version is $\Lang{max-sat}$, where the constraint
system is a propositional formula in conjunctive normal form, and
the goal is to find an assignment that satisfies the largest number of
clauses possible. The problem is $\Class{NP}$-complete even restricted
to formulas with at most two literals per clause~\cite{GareyJS76}. It
is also the canonical complete problem for the optimization class
$\Class{MaxSNP}$ and, thus, a 
central topic in the research of approximation
algorithms~\cite{PapadimitriouY91}. Many real-world problems can
be encoded as $\Lang{max-sat}$ instances, which led to the successful
development of exact solvers (see Chapter~23 and~24
in~\cite{HandbookSAT}). Following the positive example of $\Lang{sat}$
solvers, these tools became ever better over the last
decades~--~regularly breaking alleged theoretical barriers in
practice. In search of an explanation for this phenomenon,
theoreticians studied the parameterized complexity of
$\Lang{max-sat}$~\cite{AlonGKSY11,CrowstonGJRS13,DellKLMM17,IwataOY14,NarayanaswamyRRS12,ReedSV2004},
which resulted in new concepts such as \emph{parameterization above
a guarantee}~\cite{MahajanR99} or \emph{dual
parameterizations}~\cite{RazgonO09}.

With membership in the class $\Para\Class P$ (or $\Class{FPT}$) of
fixed-parameter tractable problems settled for many variants of
$\Lang{max-sat}$, a new 
question has surfaced both in theoretical and practical research over
the last decade: Which problems admit \emph{parallel} fpt-algorithms,
i.\,e., which problems lie in $\Para\Class{NC}$, the parameterized
version of~$\Class{NC}$? The vertex cover problem is the poster child
for such a problem as it lies even in $\Para\Class{AC}^0$, which is
the smallest commonly studied parameterized class and can be thought
of as ``solvable with fpt-many parallel processing units in constant
time''~\cite{BannachST15}. Many of the important tools underlying fpt-theory, such as
search trees, graph decompositions, or kernelizations, have been
adapted to the parallel setting by different research
groups~\cite{Abu-KhzamK20, BannachST15, ChenF20, PilipczukSTS2018}. 

In this paper we study the parallel complexity of maximum
satisfiability problems for various parameterizations. We show that
the parallel fpt-toolkit can be used to establish parallel algorithms
for \Lang{max-sat} parameterized by the solution size or parameterized
above some guarantee. We also develop dedicated algorithms for the
problem parameterized by the structural parameters \emph{treewidth},
\emph{feedback vertex set number}, \emph{treedepth}, and \emph{vertex
  cover number} and observe an ever higher level of achievable
parallelization.  Our most technical contribution is a parallel
algorithm for $\PLang[k]{almost-2sat}$, which is \Lang{max-2sat} for
the dual parameterization where we try to satisfy at least $m-k$
clauses in a given \textsc{2cnf} formula ($m$~is always the number of
clauses, $n$~the number of variables, $k$~a positive integer
parameter, and ``$\mathrm p$-'' indicates a parameterized problem with
the index being the parameter; variables occurring in problem names
such as in $\Lang{$d$sat}$ are fixed constants). This problem has
stubbornly resisted all known techniques in the parallel fpt-toolkit:
First, one cannot use algorithmic meta-theorems that are often used to
show membership in~$\Para\Class{NC}$. The algorithmic meta-theorems
for second-order logic~\cite{BannachT16} fail as the underlying
incidence graphs generally do not have bounded treewidth, and those
for first-order logic~\cite{ChenFH17, FlumG03, PilipczukSTS2018} fail
as the satisfiability of a \textsc{2cnf} formula is not first-order
definable. Second, the central tool for showing that it lies in
$\Para\Class P$, namely iterative
compression~\cite{RazgonO09,ReedSV2004}, is~--~as the name
suggests~--~highly \emph{sequential}.

We develop new tools that go beyond the established toolkits and involve two  ideas. First,
we make a simple, but non-trivial, observation concerning the parallel
computation of graph flows. While computing flows is
$\Class P$-complete~\cite{GoldschlagerSS82} and,
thus, most likely \emph{not} paralellizable and while even computing a
0-1-flow in parallel is a long standing open problem~\cite{KarpUW86},
we observe that computing a flow of \emph{parameter value}~$k$ can be
done in $k$ consecutive rounds of a \emph{parallel} 
Ford-Fulkerson~\cite{fordF56} step.
The second idea is more
complex, as we study a seemingly different problem: vertex cover, but not
with the solution size as the parameter, but with
the (smaller and hence less restrictive) parameter ``integrality
excess of the~\textsc{lp}.'' An fpt-reduction from
parameterized $\Lang{almost-2sat}$ to this vertex cover version is
well known~\cite{RazgonO09}. To compute vertex covers
for this ``looser'' parameter in parallel, we combine results by Iwata,
Oka and Yoshida~\cite{IwataOY14} on the properties of the Hochbaum
network underlying the linear program and apply the earlier-mentioned
observations on graph flows.
In detail, the contributions of this manuscript can be grouped as follows, see also
Table~\ref{table:overview}.

\begin{table (possibly in appendix)}[htbp]
  \caption{Variations of the maximum satisfiability problem studied
    within this paper. The lower bounds are the trivial ones, while
    the upper bounds are proven in the referenced theorems or
    lemmas. If the result is marked as constructive, a corresponding
    optimal assignment can be produced (this is either proven
    directly, or the presented algorithm can be modified in an obvious
  way). The blue headlines indicate the technique used to obtain the
  results.}
  \setlength\tabcolsep{3.25pt}
  \begin{tabular}{llllcl}\small
                                            & \hbox to 0pt{\it Complexity Bound\hss} &                               &                   &          &                                               \\ 
    \emph{Problem}                          & \emph{Lower}
                                            & \emph{Upper}                           & \emph{Clause Size}            & \emph{Construct.} & \emph{Reference}                                         \\[1ex]
    \technique{Color Coding}                &                                        &                               &                   &          &                                               \\
    $\PLang[k,t]{max-$\delta$-circuit-sat}$ & $\Para\Class{AC}^0$                    & $\Para\Class{AC}^0$           & unbounded         & \check   & Theorem~\ref{theorem:max-circuit-sat}         \\
    $\PLang[k]{max-sat}$                    & $\Para\Class{AC}^0$                    & $\Para\Class{AC}^0$           & unbounded         & \check   & Corollary~\ref{corollary:max-sat}             \\
    $\PLang[k]{max-nae-sat}$                & $\Para\Class{AC}^0$                    & $\Para\Class{AC}^0$           & unbounded         & \check   & Corollary~\ref{corollary:max-sat}             \\
    $\PLang[k,d,x]{max-exact-sat}$          & $\Para\Class{AC}^0$                    & $\Para\Class{AC}^0$           & $\leq d$          & \check   & Corollary~\ref{corollary:max-sat}             \\
    $\PLang[k,d]{max-dnf}$                  & $\Para\Class{AC}^0$                    & $\Para\Class{AC}^0$           & $\leq d$          & \check   & Corollary~\ref{corollary:max-sat}             \\
    $\PLang[g]{max-sat-above-half}$         & $\Para\Class{TC}^0$                    & $\Para\Class{TC}^0$           & unbounded         & \uncheck & Theorem~\ref{theorem:max-sat-above-guarantee} \\[1ex]
    \technique{Algebraic Techniques}        &                                        &                               &                   &          &                                               \\
    $\PLang[g]{max-e$d$sat-above-average}$  & $\Para\Class{AC}^0$                    & $\Para\Class{AC}^0$           & $= d$             & \uncheck & Lemma~\ref{lemma:max-d-sat-aa}                \\[1ex]
    \technique{Graph Flows}                 &                                        &                               &                   &          &                                               \\
    $\PLang[k]{almost-nae-2sat}$            & $\Para\Class{L}$                       & $\Para\Class{NL}^{\uparrow}$  & $\leq 2$          & \check   & Theorem~\ref{theorem:almost-2sat}             \\
    $\PLang[k]{almost-2sat}$                & $\Para\Class{NL}$                      & $\Para\Class{NL}^{\uparrow}$  & $\leq 2$          & \check   & Theorem~\ref{theorem:almost-2sat}             \\[1ex]
    \technique{Graph Extensions}            &                                        &                               &                   &          &                                               \\
    $\PLang[k]{almost-nae-sat(2)}$          & $\Para\Class{L}$                       & $\Para\Class{L}$              & unbounded         & \uncheck & Theorem~\ref{theorem:almost-sat(2)}           \\
    $\PLang[k]{almost-sat(2)}$              & $\Para\Class{L}$                       & $\Para\Class{L}$              & unbounded         & \uncheck & Theorem~\ref{theorem:almost-sat(2)}           \\[1ex]
    \technique{Reduction to Vertex Cover}   &                                        &                               &                   &          &                                               \\
    $\PLang[k]{almost-dnf}$                 & $\Para\Class{AC}^0$                    & $\Para\Class{AC}^0$           & unbounded         & \check   & Theorem~\ref{theorem:almost-dnf}              \\
    $\PLang[k]{min-sat}$                    & $\Para\Class{AC}^0$                    & $\Para\Class{AC}^0$           & unbounded         & \check   & Lemma~\ref{lemma:min-sat}                     \\[1ex]
    \technique{Dynamic Programming}         &                                        &                               &                   &          &                                               \\
    $\PLang[\mathrm{vc}]{partial-max-sat}$  & $\Para\Class{TC}^0$                    & $\Para\Class{TC}^0$           & unbounded         & \check   & Lemma~\ref{lemma:structural-vc}               \\         
    $\PLang[\mathrm{td}]{partial-max-sat}$  & $\Para\Class{TC}^0$                    & $\Para\Class{TC}^{0\uparrow}$ & unbounded         & \check   & Lemma~\ref{lemma:structure-td}                \\         
    $\PLang[\mathrm{fvs}]{partial-max-sat}$ & $\Para\Class{L}$                       & $\Para\Class{TC}^{1\uparrow}$ & unbounded         & \check   & Lemma~\ref{lemma:structure-fvs}               \\         
    $\PLang[\mathrm{tw}]{partial-max-sat}$  & $\Para\Class{L}$                       & $\Para\Class{AC}^{2\uparrow}$ & unbounded         & \check   & Lemma~\ref{lemma:structure-tw}                \\             
  \end{tabular}
  \label{table:overview}  
\end{table (possibly in appendix)}

\subparagraph*{Contribution I.}
We settle the parallel complexity of \Lang{max-sat} for the canonical
parameters $k$ (solution size) and $g$ (solution size minus
$\lceil\frac{m}{2}\rceil$): $\PLang[k]{max-sat} \in \Para\Class{AC}^0$, 
but $\PLang[g]{max-sat}$ is
$\Para\Class{TC}^0\text{-complete}$. If we assume that clauses have
size exactly~$d$ ($\Lang{max-e$d$sat}$), we show that an ``above
average version'' lies in $\Para\Class{AC}^0$ as well.

\subparagraph*{Contribution II.}
We study variants of $\PLang[k]{almost-sat}$, i.\,e., of \Lang{max-sat}
parameterized dually, and present, for the first time,
parallel algorithms for this problem on various classes of
\textsc{cnf}s. The main achievement is a $\Para\Class{NC}$ algorithm
for the problem restricted to \textsc{2cnf}s.

\subparagraph*{Contribution III.}
The structural parameters \emph{vertex cover number}, \emph{treedepth},
\emph{feedback vertex set number}, and \emph{treewidth} are partially ordered,
meaning that graphs of bounded vertex cover number have bounded
treedepth and so on. It is 
known that \Lang{max-sat} is in $\Para\Class{P}$ parameterized by any
of these, but the sequential algorithms tend to hide beneficial properties
gained by more restrictive parameterizations. We show that we 
obtain a higher level of parallelization for larger parameters
(reaching from $\Para\Class{TC}^0$ and $\Para\Class{TC}^{0\uparrow}$,
over $\Para\Class{TC}^{1\uparrow}$, up to
$\Para\Class{AC}^{2\uparrow}$). Additionally, our algorithms are
constructive (they output an optimal assignment), which is
in contrast to existing parallel meta-theorems.

As byproducts, we establish results that may be of independent interest: First, we present an
alternative characterization of the
``up-classes''. Second, we lower the complexity of the
feedback vertex set problem to $\Para\Class{L}^{\uparrow}$, which is
obtained ``by iterating a $\Para\Class{L}$ computation
parameter-many times.'' Third, we obtain $\Para\Class{NC}$ algorithms
for problems that can be reduced to $\PLang[k]{almost-2sat}$ which
includes, in particular, the odd cycle transversal problem
(can we make a given graph bipartite by deleting $k$ vertices?).

\subparagraph*{Related Work.}

The parameterized complexity of \Lang{max-sat} is an active field of
research dating back the pioneering work by Mahajan and
Raman~\cite{MahajanR99}. Since then, parameterized algorithms for ever
looser parameters have been found~\cite{CrowstonFGJRTY11,CrowstonGJY12,GutinJSY13} or their existence
has been refuted~\cite{CrowstonGJRS13}. This research has also
branched out into the study of preprocessing algorithms~\cite{GaspersS11,GaspersS14},
parameterized heuristics~\cite{Szeider11}, and algorithms utilizing structural
decompositions~\cite{DellKLMM17,Grohe06}. However, to the best of our knowledge, not yet
to parallel parameterized algorithms.

While research on parallel fixed-parameter algorithms dates back to
the early 1990s to the study of the \emph{space} complexity of
parameterized problem~\cite{CaiCDF97} (via the inclusion chain
$\Class{NC}^1 \subseteq \Class{L} \subseteq \Class{NL} \subseteq
\Class{AC}^1$), a systematic study of parallel fixed-parameter
algorithms started only in the last decade~\cite{ElberfeldST15}. Since then, a toolbox has
been compiled that contains algorithmic meta-theorems both for monadic
second-order logic~\cite{BannachT16} and for first-order
logic~\cite{ChenFH17, FlumG03, PilipczukSTS2018}.

\subparagraph*{Organization of this Paper.}

After some preliminaries in the next section, we study \Lang{max-sat}
parameterized by the solution size and parameterized above a
guarantee in Section~\ref{section:maxsat-solution}. We continue and
study \Lang{max-sat} variants with a dual parameterization in
Section~\ref{section:maxsat-dual}. The largest and technical most
involved part here is a parallel algorithm for
$\PLang[k]{almost-2sat}$. Finally, we consider structural
parameterizations of \Lang{max-sat} in Section~\ref{section:structure}
and establish a connection between the level of parallelization we can
achieve and the used parameter.

\section{Background on Parameterized Problems and Classes}
\label{section:preliminaries}

\subparagraph*{Propositional Logic and MaxSAT.}

We assume an infinite supply of \emph{propositional variables}
$x_1,x_2,\dots$ and call a variable $x$ or its negation $\neg x$ a
\emph{literal}.  A \emph{propositional formula in conjunctive normal
  form} (a \textsc{cnf}) $\phi$ is a conjunction of disjunctions of
literals, for instance
$\phi=(x_1\vee x_2\vee \neg x_2)\wedge(x_1)\wedge(x_1)\wedge(x_2\vee
x_2)$. We write $\vars(\phi)$ for the set of variables in~$\phi$ and
$\clauses(\phi)$ for the multiset of clauses, which are the sets of literals
in the disjunctions, e.\,g., $\clauses(\phi) = \bigl\{\{x_1,x_2,\neg x_2\},
\{x_1\}, \{x_1\}, \{x_2\}\bigr\}$. We denote
$\left|\vars(\phi)\right|$ by~$n$ and $\left|\clauses(\phi)\right|$
by~$m$ (so $n=2$ and $m=4$ in the example), and let $m_\emptyset$ be the number of
empty clauses.

An \emph{assignment} $\beta\colon\vars(\phi)\rightarrow\{0,1\}$ maps
every variable of $\phi$ to a truth value. It satisfies a
literal $\ell$ if $\ell=x$ and $\beta(x)=1$ or if $\ell=\neg x$ and
$\beta(x)=0$. Furthermore, it \emph{satisfies} a clause~$C$ (denoted
by $\beta\models C$) if it satisfies at least one literal in it; it
\emph{nae-satisfies} a clause if it additionally falsifies at least
one literal (``not-all-equal-satisfies'').    

The \Lang{max-sat} problem asks, given a \textsc{cnf}
$\phi$ and a number $k$, whether there is an assignment~$\beta$ that
satisfies at least $k$ clauses. If $\beta$ satisfies all $m$ clauses,
then $\beta\models\phi$, i.\,e., $\beta$ is a model of
$\phi$. Variations are obtained by modifying the
condition of a clause being satisfied, e.\,g., in \Lang{max-nae-sat}
we seek an assignment that nae-satisfies at least $k$ clauses.

\subparagraph*{Graphs, Networks, and Flows.}

In this paper, graphs are pairs $G = (V,E)$ of finite sets of
vertices and edges. In this context, $n$ denotes~$|V|$ and $m$
denotes~$|E|$. For 
undirected graphs, edges are two-element subsets of~$V$, for directed
graphs (digraphs) $E \subseteq V \times V$. A \emph{walk in~$G$ of
length~$p$} is a sequence $(v_0,\dots,v_p)$ of vertices $v_i \in V$
with $(v_i,v_{i+1}) \in E$ (or $\{v_i,v_{i+1}\} \in E$ for undirected
graphs) for all $i\in\{0,\dots, p-1\}$. A \emph{path} is a walk in
which all vertices (and hence all edges) are distinct. A \emph{cycle}
is a walk of length at least~$3$ in which all vertices are distinct
expect for the first and last, which must be identical. For a set $S
\subseteq V$ we write $G - S$ for the graph induced on the set $V
\setminus S$. For an undirected graph~$G$ the \emph{neighborhood} $N(v)$ of a
vertex~$v$ is the set $\{\,u \in V \mid \{u,v\} \in E\,\}$, the
\emph{degree} of $v$ is $|N(v)|$.

We think of digraphs $G=(V,E)$ with two designated vertices
$s,t\in V$ as \emph{networks}, and we always assume that in networks between any
two different vertices $u$ and~$v$ at most one edge is present
(either $(u,v)$ or $(v,u)$)~--~if this is not the case we may simply subdivide each edge.
A \emph{$0$-$1$-flow} from $s$ to~$t$
in $G$ is a mapping $f\colon E\rightarrow\{0,1\}$ such that for all
$v\in V\setminus\{s,t\}$ we have $\sum_{(u,v)\in E} f(u,v) =
\sum_{(v,w)\in E} f(v,w)$. The \emph{value $|f|$} of a flow is defined
as the amount $|f|=\sum_{(s,v)\in E}f(s,v)-\sum_{(w,s)\in E}f(w,s)$ of
flow leaving the source (or, equivalently, arriving at the target).
For a flow $f$ in a network $G$, the \emph{residual graph $R_f =
(V,E_f)$} contains all edges of~$G$ that are not part of the flow and
all reversed edges of the flow:
\[
  E_f = \{\,(u,v) \in E \mid f(u,v) = 0\,\} \cup \{\,(v,u) \in V \times V \mid f(u,v) = 1\,\}.
\]

\subparagraph*{Standard Parameterized Problems and Complexity Classes.}

A \emph{parameterized problem} is a set
$Q \subseteq \Sigma^* \times \mathbb N$. In an instance $(w,k)$ we
call $w$ the \emph{input} (typically a \textsc{cnf} in this paper) and
$k$ the \emph{parameter.} For instance,
$\PLang[k]{max-sat} = \{(\phi,k) \mid \phi$ has an assignment
satisfying at least $k$ clauses$\}$. We indicate the parameter as a
subscript to the leading ``p''.

A \emph{parameterized function} is a mapping $F \colon \Sigma^* \times
\mathbb N \to \Sigma^* \times \mathbb N$ such that the output
parameter is bounded in terms of the input parameter, i.\,e., there
is a function $b \colon \mathbb N \to \mathbb N$ with $k' \le b(k)$
whenever $F(w,k) = (w',k')$. The \emph{characteristic
function~$\chi_Q$} of a parameterized problem~$Q$ maps $(w,k) \in Q$ to
$(1,0)$ and $(w,k) \notin Q$ to $(0,0)$. 

In parameterized complexity theory, the class $\Para\Class P$ (also
known as $\Class{FPT}$) takes the role of $\Class P$ in classical
complexity theory. A parameterized problem $Q$ is in $\Para\Class P$
if there is an algorithm that decides whether $(w,k) \in Q$ holds in time $f(k) \cdot n^{O(1)}$ for some
computable function~$f$. A \emph{parallel} parameterized algorithm is
able to decide the same question by a
logarithmic-time-uniform\footnote{Details about uniformity will not
  be of importance in our study. We refer the interested reader
  to~\cite{BannachST15,BarringtonIS90,ChenF16} and abbreviate
  ``logarithmic-time-uniform'' with ``uniform'' in the following.}
family of unbounded fan-in circuits of depth
$O(\log^i n)$ for some fixed~$i$ (note that the depth does not depend
on~$k$) and size $f(k) \cdot n^{O(1)}$. The problem is then in the class
$\Para\Class{AC}^i$ or, in the presence of \emph{threshold gates},
$\Para\Class{TC}^i$. Define $\Para\Class{NC}$ as the union of all these
$\Para\Class{AC}^i$ classes or, equivalently, the union of all $\Para\Class{TC}^i$ classes.

\subparagraph*{Up-Classes.}
The ``up-arrow notation'' was originally introduced in the context of
parameterized circuit classes~\cite{BannachST15} to denote circuits
that arise from taking a circuit of a certain depth 
(like $\log n$) and then allow ``parameter-dependent-many
layers'' of such circuits (resulting in a depth like $f(k)
\log n $). In this paper, we define the notation as the ``closure of
a parameterized function class under parameter-dependent-many iterations of
linear functions,'' which yields the same circuit classes, but
also yields natural ``up-versions'' of $\Para\Class L$ and
$\Para\Class{NL}$. In detail, we take a parameterized function class
and allow the functions in it to be applied to an input
not just once, but rather ``parameter-dependent-many times.''
One must be a bit careful, though, to ensure that the intermediate
results do not get too large. We require that the function we
apply iteratively causes only a linear increase in the output size. For
this, let us call a parameterized function $F$ \emph{linear} if
$|F(w,k)| \le f(k) \cdot |w|$ for some computable~$f$.
\begin{definition}\label{definition:up}
  Let $\Para\Class{FC}$ be a class of parameterized functions. A
  parameterized function~$F$ lies in $\Para\Class{FC}^\uparrow$ if
  there are (1) an ``initial'' function $I \in \Para\Class{FC}$, (2) a
  linear ``iterator'' function $L \in \Para\Class{FC}$, and (3) a
  computable ``iteration number'' function $r \colon \mathbb N \to
  \mathbb N$, such that $F(w,k) = L^{r(k)}(I(w,k))$, where $L^r$ is
  the $r$-fold composition (or iteration) of~$L$ with itself. A
  problem lies in 
  $\Para\Class{C}^\uparrow$ if its characteristic function lies in
  $\Para\Class{FC}^\uparrow$.  
\end{definition}

\begin{figure}[htbp]
  \begin{tikzpicture}[
    every node/.style = {
      baseline,
      font = \small
    }
    ]
    \node (ac0) at (0,0)  {$\Para\Class{AC}^0$};
    \node (tc0) at (2,0)  {$\Para\Class{TC}^0$};
    \node (nc1) at (4,0)  {$\Para\Class{NC}^1$};
    \node (l)   at (6,0)  {$\Para\Class{L}$};
    \node (nl)  at (8,0)  {$\Para\Class{NL}$};
    \node (ac1) at (10,0)  {$\Para\Class{AC}^1$};

    \node (ac0up) at (0+1,-1)  {$\Para\Class{AC}^{0\uparrow}$};
    \node (tc0up) at (2+1,-1)  {$\Para\Class{TC}^{0\uparrow}$};
    \node (nc1up) at (4+1,-1)  {$\Para\Class{NC}^{1\uparrow}$};
    \node (lup)   at (6+1,-1)  {$\Para\Class{L}^{\uparrow}$};
    \node (nlup)  at (8+1,-1)  {$\Para\Class{NL}^{\uparrow}$};
    \node (ac1up) at (10+1,-1)  {$\Para\Class{AC}^{1\uparrow}$};
    
    \node (ac2) at (12,0) {$\Para\Class{AC}^2$};

    \graph[use existing nodes, edges = {semithick, >={[round]Stealth}}]{
      ac0 ->[densely dashed] tc0 -> nc1 -> l -> nl -> ac1 -> ac2;
      ac0 ->[densely dashed] ac0up;
      tc0 -> tc0up;
      nc1 -> nc1up;
      l -> lup;
      nl -> nlup;
      ac1 -> ac1up;
      ac0up -> tc0up -> nc1up -> lup -> nlup -> ac1up -> ac2;
    };
  \end{tikzpicture}
  \caption{Inclusions among parallel parameterized complexity
    classes within $\Para\Class{P}$. An arrow from $A$ to $B$ means $A\subseteq B$, and a
    dashed arrow indicates $A\subsetneq B$. The inclusions between the two rows follow from arguments
for the up-classes~\cite{BannachST15} and the other inclusions follow from
the standard inclusion chain $\Class{AC}^0 \subsetneq \Class{TC}^0 \subseteq\Class L
\subseteq \Class{NL} \subseteq \Class{AC}^1\subseteq \Class{AC}^2\subseteq \Class{P}$.}
  \label{figure:classes}
\end{figure}
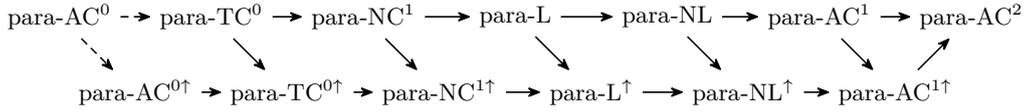

The following lemma shows that the definition is a generalization of
the original definition of~$\Para\Class{AC}^{i\uparrow}$ as the class
of problems decidable by circuits of depth $f(k)\cdot O(\log^i n)$ and
size~$f(k)\cdot n^{O(1)}$, see~\cite{BannachST15}. The lemma implies
the chain of inclusions shown in Figure~\ref{figure:classes}.

\begin{lemma (proof possibly in appendix)}\label{lemma:def-up}
  A problem $Q$ is in $\Para\Class{AC}^{i\uparrow}$ (in the sense of 
  Defintion~\ref{definition:up}) iff $Q$ can be decided by a family of
  Boolean circuits of depth $f(k)\cdot O(\log^i n)$ and size
  $f(k)\cdot n^{O(1)}$ for some computable function~$f$.
\end{lemma (proof possibly in appendix)}
\begin{proof (possibly in appendix)}[Proof of Lemma~\ref{lemma:def-up}]
  If $Q \in \Para\Class{AC}^{i\uparrow}$ via functions $I$, $L$,
  and~$f$, where $I$ is implemented by circuit family~$C_I$ and $L$
  by~$C_L$, both of depth $O(\log^i n)$, the claimed family of
  circuits simply consists of $C_I$ followed by $f(k)$ copies
  of~$C_L$.

  For the other direction, let $C = (C_{n,k})_{n,k\in\mathbb N}$ be a
  family of depth $f(k)\cdot O(\log^i n)$
  that decides~$Q$. The circuit family $C_I$ then maps an input
  $(w,k)$ to $(C',k)$ where $C'$ is the following partially evaluated
  circuit: It is  $C_{|w|,k}$ with (only) the input gates evaluated to
  the corresponding bits of~$w$. The iteration function~$L$ then takes
  a partially evaluated circuit and does $O(\log^i n)$ ``rounds of
  evaluation,'' which just means that any gate whose inputs have all
  been evaluated gets evaluated itself. Clearly, a single round of
  evaluating gates can be done by an $\Class{AC}^0$ circuit, so
  $O(\log^i n)$ rounds can be done by an $\Class{AC}^i$ circuit. Putting
  it all together, we see that after applying $C_i$ and then $f(k)$
  times the iteration function, we map the input to the fully evaluated
  circuit $C_{|w|,k}$ and can, thus, obtain the desired
  output from the output gate in the last iteration.
\end{proof (possibly in appendix)} 

The advantage of our (new, more complex) definition of up-classes is
that it naturally yields the classes $\Para\Class{L}^\uparrow$ and
$\Para\Class{NL}^\uparrow$ based on $\Para\Class{FL}$ and
$\Para\Class{FNL}$, the parameterized versions of
$\Class{FL}$ and $\Class{FNL}$. These latter classes contain all 
functions $F \colon \Sigma^* \to \Sigma^*$ such that a 
Turing machine (deterministic for $\Class L$, non-deterministic for
$\Class{NL}$) with a read-only input tape and a write-only output
tape produces~$F(w)$ on input~$w \in \Sigma^*$ using only $O(\log|w|)$
cells on its work tape (in the non-deterministic case, all halting
computations must lead to~$F(w)$ on the output tape). It is worth
noting that both $\Class{FL}$ and $\Class{FNL}$ are closed under
composition (the Immerman-Szelepcs{\'e}nyi Theorem is needed for
$\Class{FNL}$) and that they only contain functions~$F$ with $|F(w)|
\le |w|^{O(1)}$. The parameterized function classes are defined
analogously, only they contain parameterized functions $F \colon
\Sigma^* \times \mathbb N \to \Sigma^* \times 
\mathbb N$ and the machines may use $f(|w|) + O(\log |w|)$ cells on
the work tape on input $(w,k)$ for some computable function
$f\colon\mathbb N \to \mathbb N$. Note that the maximum length of
$F(w)$ is now $f'(|w|)\cdot|w|^{O(1)}$ for some other computable
function~$f'$. These classes are also closed under composition. 

To get a feeling for these classes, consider the following result on
$\PLang[k]{fvs}$:  

\begin{theorem (proof possibly in appendix)}\label{theorem:fvs}
  A size-$k$ feedback vertex set can be computed in
  $\Para\Class{FL}^{\uparrow}$, if one exists.
\end{theorem (proof possibly in appendix)}
\begin{proof (possibly in appendix)}[Proof of Theorem~\ref{theorem:fvs}]
  To prove the statement, we have to provide two functions
  $I,L\in\Para\Class{FL}$ such that $I$ maps inputs $(G,k)$ to initial
  instances $(w_0,k_0)$ and such that $L$ outputs a new instance
  $(w_{i+1},k_{i+1})$ on its output tape while reading $(w_i,k_i)$
  from its input tape. Each time the machine underlying the
  function~$L$ is run, it can freely access the output of the previous
  iteration (in a read-only fashion).

  We develop a bounded search tree algorithm of
  depth~$k$. To that end, $I$ simply translates $(G,k)$ to a list with
  a single element $\big(((G,\emptyset),k)\big)$. We will keep the
  invariant that the second part of $(G,\emptyset)$ is a partial
  feedback vertex set of $G$. The iterator function $L$ takes
  such a list, processes each item in it, and outputs a potentially
  larger list.

  In detail, for a list
  $\big(((G_1,S_1),k_1),\dots,((G_{\ell},S_{\ell}),k_{\ell})\big)$ the following
  logspace operations are performed on all tuples
  $((G_j,S_j),k_j)$: First, all degree-$1$ vertices are removed; secondly,
  all paths of degree-$2$ vertices are contracted. If
  the resulting graph $G_j'$ contains a vertex~$v$ with a self-loop,
  any solution has to contain $v$. Therefore, the machine maps
  $(G_j,k_j)$ to $((G_j'-\{v\},S_j\cup\{v\}),k_j-1)$ in the output list. Otherwise
  the minimum degree of $G_j'$ is~$3$ and a well-known fact states
  that any size-$k_j$ feedback vertex set of $G_j'$ has to contain one
  of the $3k_j$ vertices $v_1,\dots,v_{3k_j}$ of highest degree~\cite[Lemma~3.3]{CyganFKLMPPS15}. Thus,
  the machine branches on these vertices (and simulates the
  corresponding bounded search tree) by mapping $(G_j,k_j)$ to
  $\big(((G_j'-\{v_1\}, S_j\cup\{v_1\}),k_j-1),\dots,((G_j'-\{v_{3k_j}\},S_j\cup\{v_{3k_j}\}),k_j-1)\big)$.

  We can check in logspace whether one of the instances in the
  current list is a forest, in which case a solution $S$ was found. On the
  other hand, after at most $k$ iterations all parameter values fall to $0$ and, hence, the machine recognizes that it deals
  with a no-instance.

  Observe that in any iteration, the output list is larger than the
  input list by a factor of at most~$3k$ and, hence, the function $L$ is
  linear in the sense of Definition~\ref{definition:up}. 
\end{proof (possibly in appendix)}

Up-classes are closed under the
up-operator; we need and prove this only for
$\Para\Class{NL}^\uparrow$, but remark that this holds for any
well-behaved computational model.

\begin{lemma (proof possibly in appendix)}\label{lemma:closure-up}
  $\Para\Class{FNL}^{\uparrow\uparrow} = \Para\Class{FNL}^\uparrow$.
\end{lemma (proof possibly in appendix)}
\begin{proof (possibly in appendix)}[Proof of Lemma~\ref{lemma:closure-up}]
  Let $K \in \Para(\Class{FNL}^\uparrow)^\uparrow$ via $I_K, L_K \in
  \Para\Class{FNL}^\uparrow$ and $r_K$ (we call these the ``outer'' functions
  and bound); and let $I_K
  \in \Para\Class{FNL}^\uparrow$ via $I_{I_K}$, $L_{I_K}$, and
  $r_{I_K}$, and let further
  $L_K \in \Para\Class{FNL}^\uparrow$ via $I_{L_K}$, 
  $L_{L_K}$, and~$r_{L_K}$ (we call these the ``inner'' functions and
  bounds). To show $K \in \Para\Class{FNL}^\uparrow$, we must 
  specify $I'_K \in \Para\Class{FNL}$, $L'_K \in \Para\Class{FNL}$,
  and $r'_K$ that show this. We sketch the construction in the
  following, but leave out most technical details.

  The idea is as follows (we ignore the initial
  function $I_K$ for the moment, assume that it is the identity): One
  application of the outer iteration function $L_K$ consist of a 
  single application of $I_{L_K}$ 
  followed by parameter-dependent-many iterations of $L_{L_K}$. This means that
  parameter-dependent-many iterations of~$L_K$ correspond to the repeated
  application of a function $L'_K$ that sometimes applies $I_{L_K}$
  and most of the time $L_{L_K}$. To decide which function must be
  applied, we  keep track of counters $o$ and $i$ as part of
  the input and output, which count the current ``outer''
  iteration~$o$ and the current ``inner'' iteration~$i$, starting with
  $o=1$ and $i=0$. For $i=0$ the
  function $L'_K$ applies $I_{L_K}$, for $i > 0$ it applies
  $L_{L_K}$. In either case, $i$ is incremented by $1$ in the output,
  unless it exceeds $r_{L_k}$ (applied to the current input). If this
  happens, we reset $i$ to $0$ in the output and increase $o$
  by~$1$. At the very end, when $o$ has reached $r_K$, we strip the
  counters from the output, which means that we output the correct
  final value.

  There remains one technical problem with the above idea: The function
  $I_{L_K}$, which is repeatedly applied by the just-described
  function $L'_K$ to the current input, is not linear -- meaning that
  $L'_K$ will not be linear. However, while
  $I_{L_K}$ alone need not be linear, $L_K$ as a whole is (by
  definition). For this reason, our initial function $I'_K$ will map
  the input $w$ to a larger input padded by as many blank symbols as
  the maximal number of symbols that $I_K$ followed by $r_K$ many
  applications of $L_K$ could possibly have (this number is of the
  form $f(k)\cdot n^{c}$ for some computable $f$ and constant~$c$), and each iteration
  of~$L'_K$ strips its input of trailing blanks, then applies
  $I_{L_K}$ or $L_{L_K}$ as described above, and then once more adds
  as many blanks as needed so that the output has the same length as
  the input had. This clearly ensures that $L'_K$ is now a linear
  function.

  It remains to argue that the construction can be adapted to the case
  where $I_K$ is not the identity. In this case we start the
  iteration of $L'_K$ for the initial counters $o=0$ and $i=0$ and
  treat $o=0$ in a special way: For this counter, $L'_K$ applies
  $I_{I_K}$ for $i=0$ (rather than $I_{L_k}$) and $L_{I_K}$ for $i>0$
  (rather than $L_{L_K}$). The construction remains unchanged, otherwise.
\end{proof (possibly in appendix)}

\section{MaxSAT Variants Parameterized by Solution Size}
\label{section:maxsat-solution}

A natural parameterization of a problem such as \Lang{max-sat} is to
take as parameter~$k$ the size of the sought solution. It is
well-known that the corresponding problem 
$\PLang[k]{max-sat}$ is in $\Para\Class{P}$~\cite{MahajanR99}. We
prove in Section~\ref{section:maxsat-variants} that the problem lies
in $\Para\Class{AC}^0$ and that this result generalizes to a broader
range of problems. It is also known that a version with less restrictive parameter
is in $\Para\Class{P}$ as well~\cite{MahajanR99}: 
$\PLang[g]{max-sat-above-half}$ asks whether there is an
assignment that satisfies at least
$\lceil\frac{m-m_\emptyset}{2}\rceil+g$ 
clauses, where $m$ is the total number of clauses and $m_\emptyset$ the number
of empty clauses in the input. We show in
Section~\ref{section:maxsat-above-half} that this problem is strictly
harder than $\PLang[k]{max-sat}$, as it is complete for $\Para\Class{TC}^0$.

\subsection{Maximum Bounded-Circuit Satisfiability}
\label{section:maxsat-variants}

We consider four variants of $\Lang{max-sat}$, where we maximize
the number of clauses
\begin{itemize}
\item for $\PLang[k]{max-sat}$ in which at least one literal is true;
\item for $\PLang[k]{max-nae-sat}$ in which at least one literal is
  true and one is false;
\item for $\PLang[k,d,x]{max-exact-sat}$ in which exactly $x$ of the
  $d$ literals are true;
\item for $\PLang[k,d]{max-dnf}$ in which all of the $d$ literals are
  true.
\end{itemize}
All of these problems are special cases of
Problem~\ref{prob:max-delta-circuit-sat} below. For its definition, we
say that a Boolean function $f\colon\{0,1\}^n\rightarrow\{0,1\}$ is
\emph{t-robust} if for every point $x\in\{0,1\}^n$ with $f(x)=1$ there
is a set of at most $t$ indices such that $f(y)=1$ for any
$y\in\{0,1\}^n$ that equals $x$ on these indices. For instance, a
clause on $d$ literals is 1-robust, while a \emph{term} (a conjunction
of literals) is $d$-robust. We are interested in the promise
problem defined on the next page.

\clearpage
\begin{problem}[{$\PLang[k,t]{max-$\delta$-circuit-sat}$}]\label{prob:max-delta-circuit-sat}
  \begin{parameterizedproblem}
    \instance Integers $k$ and $t$, $\Class{AC}$-circuits $C_1$,
    $\dots$, $C_m$, all connected to
    the same $n$ input variables $x_1$, $\dots$, $x_n$, all with a
    single output gate, and all of depth at most~$\delta$.
    \parameter $k+t$
    \question Is there an assignment from the input variables to
    $\{0,1\}$ such that at least $k$ circuits evaluate to $1$?
    \promise All circuits compute a $t$-robust function.
  \end{parameterizedproblem}
\end{problem}

\begin{theorem (proof possibly in appendix)}\label{theorem:max-circuit-sat}
  $\PLang[k,t]{max-$\delta$-circuit-sat}\in\Para\Class{AC}^0$.
\end{theorem (proof possibly in appendix)}
\begin{proof (possibly in appendix)}[Proof of Theorem~\ref{theorem:max-circuit-sat}]{
  We start with some observations. First, a $\Para\Class{AC}^0$
  circuit can, given an assignment to the input variables, evaluate
  every input circuit $C_1,\dots,C_n$ as these are all of constant
  depth at most~$\delta$. Second, a $\Para\Class{AC}^0$ circuit can
  also check whether at least $k$ of these circuits evaluate to $1$
  (since $\Para\Class{AC}^0$ circuits can simulate threshold gates with
  parameter bounded thresholds~\cite{BannachST15}). Third, since it is
  promised that all circuits are $t$-robust, we need to
  set at most $t$ variables correctly in order to let a circuit
  evaluate to $1$. Hence, if there is a solution that satisfies $k$
  circuits, we have to find the correct truth value for at most $tk$ variables~--~of course, we do not know of which variables.

  To find them, we use the well-known color coding technique~\cite{AlonYZ95} and its
  constant-detph derandomization~\cite{BannachST15}. The proof hinges on universal
  coloring families and the fact that we can
  compute them quickly in parallel:
  \begin{definition}[Universal Coloring Families]\label{definition:universal-coloring}
    For natural numbers $n$, $k$, and $c$, an \emph{$(n,k,c)$-\penalty0universal
      coloring family} is a set $\Lambda$ of functions
    $\lambda\colon\{1,\dots,n\}\to\{1,\dots,c\}$ such that for every
    subset $S\subseteq\{1,\dots,n\}$ of size $|S| = k$ and for every 
    mapping $\mu\colon S\to\{1,\dots,c\}$ there is at least one function
    $\lambda \in \Lambda$ with $\forall s \in S\colon \mu(s) =
    \lambda(s)$. 
  \end{definition}
  \begin{fact}[Theorem~3.2 in \cite{BannachST15}]\label{fact:color-coding}
    {There is a uniform family
      $(C_{n,k,c})_{n,k,c\in\mathbb N}$ of $\Class{AC}$-circuits without
      inputs such that each $C_{n,k,c}$
      \begin{enumerate}
      \item outputs an $(n,k,c)$-universal coloring family (coded as a
        sequence of function tables), 
      \item has constant depth (independent of $n$, $k$, or $c$), and
      \item has size at most $O(n\log c \cdot c^{k^2} \cdot k^4\log^2 n)$.
      \end{enumerate}}
  \end{fact}

  To solve $\PLang[k,t]{max-$\delta$-circuit-sat}$, we use a
  $(n,tk,2)$-universal coloring family $\Lambda$ using
  Fact~\ref{fact:color-coding}. Intuitively, we ``color'' the
  variables with two colors, which we interpret as assigning truth
  values to them. Clearly, if the input is a no-instance, no coloring
  will satisfy $k$ circuits and we can correctly reject. On the other
  hand, assume there is some assignment that satisfies at least~$k$
  circuits. Then, by the above observations, there are at most $tk$
  variables $y_1,\dots,y_{tk}$ that, if set correctly, satisfy the same
  $k$ circuits. By Definition~\ref{definition:universal-coloring},
  there is at least one $\lambda\in\Lambda$ that realises exactly this
  correct assignment and, hence, by testing all colorings of~$\Lambda$
  in parallel, we can decide whether at least one assignment satisfies
  $k$ or more circuits.
}\end{proof (possibly in appendix)}

\begin{corollary}\label{corollary:max-sat}
  The problems  $\PLang[k]{max-sat}$,  $\PLang[k]{max-nae-sat}$,
  $\PLang[k,d,x]{max-exact-sat}$, and $\PLang[k,d]{max-dnf}$ are in
  $\Para\Class{AC}^0$. 
\end{corollary}

\subsection{Maximum Satisfiability Above Guarantee}
\label{section:maxsat-above-half}

The solution size is a very restrictive parameter for problems such
as \Lang{max-sat}, because \emph{every} instance has \emph{relatively
  large} solutions. In particular, let $\phi$ be a \textsc{cnf} with
$m$ clauses of which $m_\emptyset$ are empty. Then $\phi$ \emph{always} has an
assignment that satisfies at least $\lceil\frac{m-m_\emptyset}{2}\rceil$
clauses: Pick an arbitrary assignment $\beta$ and observe that either
$\beta$ or its bitwise complement satisfies half of the
clauses~\cite{MahajanR99}. Hence, $\PLang[k]{max-sat}$ is only interesting for large $k$
and to obtain efficient parallel algorithms, we require a smaller
parameterizations.

We start with a problem of the form
$Q = \{(w,k) \mid \operatorname{opt}(w) \le k\}$, where
$\operatorname{opt}(w)$ is some property to be evaluated. The new
problem has the form $Q' = \bigl\{\bigl((w,\pi),g\bigr) \mid \pi$ is an easily
checkable proof for $\operatorname{opt}(w) \ge \gamma(\pi),$ and
$\operatorname{opt}(w) \le \gamma(\pi) + g\bigr\}$.  Here, $\gamma(\pi)$
is called the \emph{guaranteed lower bound proved by~$\pi$} or just
the \emph{guarantee.}  For $Q = \PLang[k]{max-sat}$ the situation is
particularly easy, we can take as proof~$\pi$ a tautology (since there
is nothing to prove in this case) and set
$\gamma(\pi)=\lceil\frac{m-m_\emptyset}{2}\rceil$.
Note that $Q'$ is conceptionally harder than~$Q$: An
fpt-algorithm for~$Q'$ must find a (possibly large) optimal
solution, but may only use time $f(g)\cdot n^{O(1)}$ for a (possibly small)
difference~$g$.

\begin{problem}[{$\PLang[g]{max-sat-above-half}$}]\label{problem:max-sat-above}
\begin{parameterizedproblem}
  \instance A \textsc{cnf} $\phi$ with $m$ clauses of which $m_\emptyset$ are
  empty, and a difference $g \in \mathbb N$.
  \parameter $g$
  \question Is there an assignment that satisfies at least $\lceil\frac{m-m_\emptyset}{2}\rceil+g$ clauses?
\end{parameterizedproblem}
\end{problem}

Algorithms for above-guarantee parameterizations have led to a number
of algorithmic breakthroughs, for instance in the design of
algorithms for 
$\Lang{almost-2sat}$~\cite{NarayanaswamyRRS12}, linear-time
fpt-algorithms~\cite{IwataOY14}, or stricter parameterizations of
$\Lang{vertex-cover}$~\cite{GargP16}. One of these breakthroughs was
$\PLang[g]{max-sat-above-half}\in\Para\Class{P}$~\cite{MahajanR99}. The
following theorem sharpens this result by placing
$\PLang[g]{max-sat-above-half}$ in $\Para\Class{TC}^0$. This also
pinpoints the intuition that above-guarantee parameterizations are
conceptionally harder than their standard counterparts, as we obtain that
$\PLang[g]{max-sat-above-half}$ is strictly harder than
$\PLang[k]{max-sat}$ (since $\Para\Class{AC}^0\subsetneq\Para\Class{TC}^0$).

\begin{theorem (proof possibly in appendix)}\label{theorem:max-sat-above-guarantee}
  $\PLang[g]{max-sat-above-half}$ is $\leq_{\mathrm{tt}}^{\Para\Class{AC}^0}$-complete for $\Para\Class{TC}^0$.
\end{theorem (proof possibly in appendix)}
\begin{proof (possibly in appendix)}[Proof of Theorem~\ref{theorem:max-sat-above-guarantee}]{
  We prove containment with a parallel version of the
  algorithm from~\cite{MahajanR99}. The following reduction rules are
  easily seen to be safe (empty clauses cannot be satisfied and do not
  count towards the lower bound; any assignment satisfies exactly one of the
  two unit clauses):
  \begin{reductionrule}[Empty Clauses]\label{rule:maxsat-above-guarantee-empty}
    If there are empty clauses, remove them.
  \end{reductionrule}
  \begin{reductionrule}[Unit Pair]\label{rule:maxsat-above-guarantee-unit}
    If there are two unit clauses $(x)$ and $(\neg x)$, remove both.
  \end{reductionrule}
  An exhaustive application of
  Rule~\ref{rule:maxsat-above-guarantee-unit} can be carried out in
  $\Class{TC}^0$ by counting for every variable $x\in\vars(\phi)$ the
  number of unit clauses that contain $x$ or $\neg x$, respectively.

  \begin{reductionrule}[Trivial Decision]\label{rule:maxsat-above-guarantee-trivial}
    Reduce to a trivial yes-instance if there are at least $\lceil
    m/2\rceil+k$ unit clauses or at least $4k+4$ non-unit clauses.
  \end{reductionrule}
  
  Rule~\ref{rule:maxsat-above-guarantee-trivial} is safe if
  rules~\ref{rule:maxsat-above-guarantee-empty} and~\ref{rule:maxsat-above-guarantee-unit} cannot be applied:
  The amount of unit clauses alone would constitute a valid
  solution and every \textsc{cnf} on $m$ clauses with
  $m_\emptyset=0$ and $p$ non-unit clauses has an
  assignment that satisfies at least $\lceil m/2\rceil+p/4-1$
  clauses~\cite[Proposition~8]{MahajanR99}. Hence, every such formula
  with at least $4k+4$ non-unit clauses is a
  yes-instance.

  Finally, assume we have a formula $\phi$ with $m$ clauses and
  parameter $k$, to which the
  rules~\ref{rule:maxsat-above-guarantee-empty}--\ref{rule:maxsat-above-guarantee-trivial}
  cannot be applied. Then
  $m\leq(\lceil m/2\rceil+k-1)+(4k+3)=\lceil m/2\rceil+5k+2$ and,
  thus, $\lfloor m/2\rfloor\leq 5k+2$. Hence, the problem has reduced
  to the question whether there is an assignment that satisfies at least
  $6k+3$ clauses, which we can answer with
  Corollary~\ref{corollary:max-sat}.

  For hardness we perform a truth-table reduction from a
  parameterized version of the majority problem:
  $\PLang[0]{majority}$ (the majority problem ask whether a binary
  string contains more $1$s than $0$s; the trivial parameter does
  nothing). In truth-table reductions we are allowed to produce
  polynomial many instances of the target problem, query an oracle to
  solve them all at once, and then build a Boolean combination of the
  results.

  Given an instance $w=b_1\dots b_n$ of $\PLang[0]{majority}$, we
  build a formula $\phi_0=\bigwedge_{i=1}^n C_i$ with:
  \[
    C_i=\begin{cases}
      (x)      & \text{if $b_i=1$,}\\
      (\neg x) & \text{else.}
    \end{cases}
  \]
  From $\phi$ we build $n+1$ instances of
  $\PLang[g]{max-sat-above-half}$: set $g=1$ and obtain $\phi_1,\dots,\phi_n$
  from $\phi_0$ as follows:
  \begin{align*}
    \phi_i = \phi_0\wedge\bigwedge_{j=1}^i (x).
  \end{align*}

  Observe that $(\phi_0,1)\not\in\PLang[g]{max-sat-above-half}$
  iff $w$ contains the same amount of $0$s as $1$s. Then observe that,
  if $(w,0)\in\PLang[0]{majority}$, we have for \emph{all}
  $i\in\{0,\dots,n\}$ that
  $(\phi_i,1)\in\PLang[g]{max-sat-above-half}$. On the other
  hand, if $(w,0)\not\in\PLang[0]{majority}$, then there is an index
  $\ell\in\{0,\dots,n\}$ such that
  $(\phi_0,1),\dots,(\phi_{\ell-1},1)$ are instances of
  $\PLang[g]{max-sat-above-half}$; but $(\phi_{\ell},1)$ is
  \emph{not}. We conclude:
  \[
    \PLang[0]{majority}\leq_{\mathrm{tt}}^{\Para\Class{AC}^0}\PLang[g]{max-sat-above-half}.\qedhere
  \]  
}\end{proof (possibly in appendix)}

This result is also tight in the sense that relaxing the parameterization
further leads to an intractable problem: Let $r_1,\dots,r_m$ be the
number of literals in the clauses of a \textsc{cnf} $\phi$, then
$E(\phi) :=\sum_{i=1}^m(1-2^{-r_i})$ is the expected
number of clauses satisfied 
by a random truth assignment. It is well-known that an assignment that
satisfies at least $E(\phi)$ clauses can be found in
polynomial time~\cite{CrowstonGJRS13}. However, the problem
$\PLang[g]{max-sat-above-average}$, which asks whether we can satisfy
\emph{at least $E(\phi)+g$ clauses,} is intractable:
\begin{fact}[\cite{CrowstonGJRS13}]\label{fact:max-sat-aa}
  $\PLang[g]{max-sat-above-average}$ is $\Para\Class{NP}$-complete.
\end{fact}

This result requires clauses of arbitrary
size. If all clauses contain \emph{exactly}~$d$
distinct and non-complementary literals, the problem becomes fixed-parameter
tractable~\cite{AlonGKSY11}. Note that $E(\phi) = (1-2^d)m$ holds
in this case. The corresponding algorithm is
quite simple and can directly be parallelized (however, it requires
non-trivial results about algebraic representations of formulas that
were proven in~\cite{AlonGKSY11}; see also Section~9.2
in~\cite{CyganFKLMPPS15} for details).
\begin{lemma (proof possibly in appendix)}\label{lemma:max-d-sat-aa}
  $\PLang[g]{max-e$d$sat-above-average}\in\Para\Class{AC}^0$.
\end{lemma (proof possibly in appendix)}
\begin{proof (possibly in appendix)}[Proof of
    Lemma~\ref{lemma:max-d-sat-aa}]{ 
  Let $\phi$ with $\vars(\phi)=\{x_1,\dots,x_n\}$ and
  $\clauses(\phi)=\{C_1,\dots,C_m\}$ be the input, where each $C_i$
  contains exactly $d$ distinct and non-complementary literals, and let us denote
  by $\vars(C_i)$ the variables that occur as literals in~$C_i$ (i.\,e., $|\vars(\phi)|=d$). We
  identify the truth values \texttt{false} and \texttt{true} (which we
  usually identify with $0$ and $1$) with $-1$ and $1$, and we consider
  the following polynomial:
  \[
    X(x_1,\dots,x_n) = \sum_{1\leq i\leq m}\Bigl(
    1-\textstyle\prod_{x_j\in
      \vars(C_i)}\bigl(1+\mathrm{sign}(x_j,C_i)x_j\bigr) 
    \Bigr),
    \]
  where $\mathrm{sign}(x_j,C_i)=-1$ if $x_j\in C_i$ and
  $\mathrm{sign}(x_j,C_i)=1$ if $\neg x_j\in C_i$. As observed in
  \cite{AlonGKSY11}, each product $\prod_{x_j\in
    \vars(C_i)}\bigl(1+\mathrm{sign}(x_j,C_i)x_j\bigr)$ equals
  $2^d$ if $C_i$ is falsified by an assignment $x_1,\dots,x_n$
  and equals $0$ when it is satisfied. This in turn means that every
  satisfied clause contributes $1$ towards the sum in
  $X(x_1,\dots,x_n)$, while each falsified clause contributes
  $1-2^d$. Thus, $X(x_1,\dots,x_n) = m - 2^d(m - s) = 2^d (s -
  (1-2^{-d})m) = 2^d(s - E(\phi))$ where $s$ is  
  the number of clauses satisfied by $x_1,\dots,x_n$ and $E(\phi)$ is
  the expected number of satisfies clauses. Thus, $X
  (x_1,\dots,x_n) \ge g \cdot 2^d$ iff $x_1,\dots,x_n$ is an
  assignment that satisfies $g$ clauses more than the expected number
  of clauses satisfied by a random assignment. Note that in $g\cdot
  2^d$ the number $g$ is the parameter and $d$ is a constant.

  Since the size of 
  the clauses is a fixed constant $d$, we can write $X$ explicitly as
  sum of at most $(2^d+1)m$ monomials. The list of these monomials can
  be produced on input $\phi$ by a $\Para\Class{FAC}^0$ circuit.

  We now apply the following reduction rule that follows directly from
  the results in~\cite{AlonGKSY11}:
  \begin{reductionrule}[see Lemma~9.19 and Lemma~9.12 in~\cite{CyganFKLMPPS15}]
    If there are at least $4\cdot 9^d\cdot 4^d\cdot g^2$ monomials,
    reduce to a trivial yes-instance.
  \end{reductionrule}

  Since $d$ is a constant and $g$ a parameter, a $\Para\Class{AC}^0$
  circuit can check whether or not the rule can be applied. If it
  applies, we are done. Otherwise at most $O(g^2)$ variables
  appear in the monomial representation of $X$ and we can solve the
  problem via brute force.  
}\end{proof (possibly in appendix)}

\section{Dual Parameterizations for Variants of MaxSAT}
\label{section:maxsat-dual}

We saw that \Lang{max-sat} can be solved in parallel
when parameterized by the solution size. However, since
\Lang{max-sat} instances always only have large solutions, we moved on
to seeking solutions of size $\lceil\frac{m-m_\emptyset}{2}\rceil + g$ and then of
size $E(\phi) + g$ for parameter~$g$. We saw that the complexity increases, but also
that  parallel parameterized algorithms are still possible for most
variants. Now, we consider \emph{dual parameterizations} where the
sought solution size is~$m-k$. The corresponding problem is
called $\PLang[k]{almost-sat}$ or, if the input formula comes from a
family $\Phi$, $\PLang[k]{almost-$\Phi$}$. These problems are even harder
and in order to solve them, we must, in particular, be able to decide
$\Phi$ for inputs with $k=0$:
\begin{observation}\label{observation:almost-lowerbound}
  If $\Phi$ is a family of propositional formulas such that deciding
  satisfiability for $\Phi$ is hard for a complexity class
  $\mathcal{C}$, then $\PLang[k]{almost-$\Phi$}$ is hard for
  $\Para\mathcal{C}$.
\end{observation}

Hence we have that $\PLang[k]{almost-3sat}$ is $\Para\Class{NP}$-hard,
$\PLang[k]{almost-horn}$ is $\Para\Class{P}$-hard, and
$\PLang[k]{almost-2sat}$ is $\Para\Class{NL}$-hard. However, the
observation does \emph{not} provide any hint on upper bounds, e.\,g.,
it is not clear whether
$\PLang[k]{almost-2sat}\in\Para\Class{NL}$. Since we are interested in
parallel algorithms, we study families
of formulas that can be decided in subclasses of $\Class{P}$:
$\PLang[k]{almost-nae-2sat}$ and $\PLang[k]{almost-2sat}$ in
Section~\ref{section:dual-krom} ($\Lang{nae-2sat}\in\Class{L}$ and
$\Lang{2sat}\in\Class{NL}$), $\PLang[k]{almost-nae-sat(2)}$ and
$\PLang[k]{almost-sat(2)}$ in Section~\ref{section:dual-twice}
($\Lang{nae-sat(2)}\in\Class{L}$ and $\Lang{sat(2)}\in\Class{L}$), and
$\PLang[k]{almost-dnf}$ in Section~\ref{section:dual-dnf}
($\Lang{dnf}\in\Class{AC}^0$).

\subsection{Dual Parameterization for Krom Formulas}
\label{section:dual-krom}

Our first result about dual parameterizations is the technically
most involved part of the paper. In this section we prove the
following theorem:

\begin{theorem}\label{theorem:almost-2sat}
$\PLang[k]{almost-nae-2sat}$ and $\PLang[k]{almost-2sat}$ both lie in
$\Para\Class{NL}^{\uparrow}$. 
\end{theorem}

The proof of the theorem is based on the well-known equivalence
between $\PLang[k]{almost-2sat}$ and another member of the family
of above-guarantee problems (see Section~\ref{section:maxsat-above-half}):

\begin{problem}[{$\PLang[g]{vc-above-matching}$}]\label{problem:vc-m}
\begin{parameterizedproblem}
  \instance A graph $G=(V,E)$, a matching $M \subseteq E$, a difference $g
  \in \mathbb N$.
  \parameter $g$
  \question Is there a set $S\subseteq V$ with $|S|\leq |M|+g$ and
  $e\cap S\neq\emptyset$ for every $e\in E$?
\end{parameterizedproblem}
\end{problem}

While it is known that $\PLang[k]{vertex-cover}\in\Para\Class{AC}^0$
(Theorem 4.5 in~\cite{BannachST15}), we will need the rest of this
section to prove the following theorem:

\begin{theorem}\label{theorem:vc-above-matching}
  $\PLang[g]{vc-above-matching} \in \Para\Class{NL}^\uparrow$.
\end{theorem}

Theorem~\ref{theorem:almost-2sat} follows directly with the following
lemma, which shows that the required 
well-known reductions~\cite{CyganPPW13,NarayanaswamyRRS12} can, firstly, be implemented in
$\Para\Class{FAC}^0$ and, secondly, the last reduction can also compute
the necessary matching as part of its output.

\begin{lemma (proof possibly in appendix)}\label{lemma:reductions}
  $$
  \PLang[k]{almost-nae-2sat} \reduce
  \PLang[k]{almost-2sat} \reduce
  \PLang[g]{vc-above-matching}.
  $$
\end{lemma (proof possibly in appendix)}
\begin{proof (possibly in appendix)}[Proof of Lemma~\ref{lemma:reductions}]
  Let $(\phi,k)$ be the input of
  $\PLang[k]{almost-nae-2sat}$. We generate a new formula $\psi$ by
  replacing every clause $(\ell\vee\ell')\in\clauses(\phi)$ with
  $(\ell\vee\ell')\wedge(\neg\ell\vee\neg\ell')$. Clearly, $\psi$ is
  satisfiable iff $\phi$ has an assignment that makes exactly one
  literal in every clause true (i.\,e., if $\phi$ has an
  nae-assignment). Since any assignment satisfies at
  least one of $(\ell\vee\ell')$ and $(\neg\ell\vee\neg\ell')$,
  deleting a clause in $\phi$ is equivalent to deleting a
  clause in $\psi$. Thus:
  \[
    (\phi,k)\in\PLang[k]{almost-nae-2sat}
    \Leftrightarrow
    (\psi,k)\in\PLang[k]{almost-2sat}.
  \]
     
  For the next reduction, we first establish the following: 
  \[
    \PLang[k]{almost-2sat} \reduce \PLang[k]{variable-deletion-2sat},
  \]
  where the latter problem contains all pairs $(\phi, k)$ such that
  $\phi$ is a \textsc{2cnf} formula in which we can delete $k$
  variables together with all clauses containing them in order to make
  $\phi$ satisfiable. We replace each variable~$x\in\vars(\phi)$ with copies $x_1,\dots,x_m$
  such that each copy occurs in exactly one clause, i.\,e., if clause
  $C_i$ originally contains variable~$x$, it will contain $x_i$ in the
  new formula.  We add equality constraints to ensure that all copies obtain the same value:
  \[
    \bigwedge_{x\in\vars(\phi)}\bigwedge_{i=1}^{m}\bigwedge_{j\neq  i}
    \big(
    (\neg x_i\vee x_j)
    \wedge
    (x_i\vee\neg x_j)
    \big).
  \]
  It is easy to see that the
  resulting formula is satisfiable iff $\phi$ is
  satisfiable. Furthermore, deleting a variable $x_i$ has exactly the
  same effect as deleting the clause $C_i$ from $\phi$.

  Finally, we show
  $\PLang[k]{variable-deletion-2sat} \reduce
  \PLang[g]{vc-above-matching}$. Let again $(\phi,k)$ be the input. We
  construct an undirected graph that contains for every variable~$x$
  two vertices $x^+$ and $x^-$ that are connected by an
  edge. Furthermore, every clause (recall that these are binary) is
  represented by an edge between the vertices of the corresponding
  literals.  The resulting graph has a perfect matching, namely
  $M=\{\,\{x^+,x^-\}\mid x\in \vars(\phi)\,\}$. Thus, if $\phi$ has
  $n$ variables, any vertex cover in $G$ needs to have size at least
  $n$. In 
  fact, if $\phi$ is satisfiable, there will be a vertex cover of size
  $n$ (the satisfying assignment). Deleting a variable $x$ and all
  clauses containing $x$ from $\phi$ is equivalent to selecting both,
  $x^+$ and $x^-$, to the vertex cover. Hence, $\phi$ can be made
  satisfiable by deleting at most $k$ variables iff $G$ contains a vertex
  cover of size $n+k$. Note that we do not have to compute the perfect
  matching but rather obtain it directly from the construction and,
  hence, this shows that we can map $(\phi,k)$ to $((G,M),k)$.

  The whole reduction chain can be carried out by a
  $\Para\Class{FAC}^0$ function: We preserve the parameter ($k$ is
  always mapped directly to~$k$) and perform otherwise only simple
  projections. In fact, we only rename variables and add some fixed
  additional clauses.  
\end{proof (possibly in appendix)}

\subsubsection{A Parallel Algorithm to Compute 0-1-Flows}
\label{section:cuts}

Our algorithm behind Theorem~\ref{theorem:vc-above-matching} will
heavily rely on repeated flow computations. Maximum flows can be computed in
polynomial time with, say, the  Ford--Fulkerson
algorithm~\cite{fordF56}. However, computing the value of a
weighted maximum flow is $\Class{P}$-complete~\cite{GoldschlagerSS82},
and whether we can compute a 0-1-flow in parallel is a long
standing open problem~\cite{KarpUW86}. It is worth noting that a
maximum 0-1-flow can be computed in randomized 
$\Class{NC}$ via a reduction to the maximum matching problem in
bipartite graphs~\cite{KarpUW86}. Unfortunately, this reduction is
\emph{not} parameter-preserving and, thus, we may not apply 
parameterized matching algorithms~\cite{BannachT18}.

Our objective in this section is to show that a flow of
value~$k$ can be computed in parallel; more precisely, that there is a
function in $\Para\Class{FNL}^\uparrow$ mapping $((G,s,t),k)$ to a 0-1-flow
of value~$k$ from $s$ to~$t$, if it exists, and otherwise to a maximum
flow (formally, the output of a parameterized function must be a pair
where the second component is a new parameter value, but we will not
need this here and just silently assume that this value is set to,
say,~0).

\subparagraph*{Computing Paths in FNL.}

It is well-known that the reachability problem in digraphs is the canonical complete
problem for $\Class{NL}$ and, thus, it may seem trivial that we should
be able to compute paths in $\Class{FNL}$. However, being
able to tell whether there is a path form $s$ to~$t$ is not the same
as actually finding such a path: For instance, it is known
that in tournaments (digraphs with exactly one 
edge between any pair of vertices) reachability lies in $\Class{AC}^0$, the
distance problem is $\Class{NL}$-complete, and constructing a path
longer than the shortest path by a factor of $1+\epsilon$ can be done
in deterministic logarithmic space~\cite{NickelsenT05} -- meaning that
reachability and path construction can have vastly different
complexities. Nevertheless:

\begin{lemma (proof possibly in appendix)}\label{lemma:path-nl}
  There is a function in $\Class{FNL}$ that maps $(G,s,t)$ to a
  shortest path from $s$ to~$t$, provided it exists.
\end{lemma (proof possibly in appendix)}

\begin{proof (possibly in appendix)}[Proof of Lemma~\ref{lemma:path-nl}]
  Let $(G,s,t)$ with $G = (V,E)$ be given as input.   Since the
  distance problem is complete for both $\Class{NL}$ and
  $\Class{coNL}$, an $\Class{NL}$-machine can compute
  the distance $d$ from a given vertex $v\in V$ to~$t$
  in~$G$. Furthermore, if $d < \infty$, the machine can also compute
  all vertices $u \in V$ that are one step nearer to~$t$, i.\,e.,
  that have distance $d-1$. Finally, for each $v \in V$ it can chose
  one such~$u$ (say, the lexicographical smallest) and form a graph $H=(V,E')$ where
  $E'$ contains all these edges $(v,u)$. Then $H$ is a
  forest with out-degree at most~$1$ and with a unique $s$-$t$-path
  (if one exists in $G$). The machine may deterministically traverse
  and output this path. Note that the result is independent of the nondeterministic
  choices that were made during the computation.
\end{proof (possibly in appendix)}

\subparagraph*{Computing 0-1-Flows in para-FNL$^{\uparrow}$.}

The most important operation in the Ford--Fulkerson algorithm is
the computation of an \emph{augmenting path.}
An iterated application of Lemma~\ref{lemma:path-nl} therefore allows
us to compute a small flow:

\begin{theorem (proof possibly in appendix)}\label{theorem:flow}
  There is a parameterized function in $\Para\Class{FNL}^\uparrow$
  that maps $((G,s,t),k)$ to a flow from $s$ to~$t$
  in~$G$ of value~$k$, if it exists, or to a maximum flow otherwise. 
\end{theorem (proof possibly in appendix)}

\begin{proof (possibly in appendix)}[Proof of Theorem~\ref{theorem:flow}]
  To show that a function lies in $\Para\Class{FNL}^\uparrow$, we need
  to specify the initial function, the iteration function, and the
  number of iterations, see Definition~\ref{definition:up}.

  The initial function does very little: It just maps $((G,s,t),k)$ to
  $((G,s,t,f_0),k)$ where $f_0$ is the empty flow ($f_0(e) = 0$ for
  all $e \in E)$.  
  The interesting part is the iteration function in $\Para\Class{FNL}$
  (actually, it will even lie in $\Class{FNL}$), which implements a
  single step of the Ford--Fulkerson algorithm: It gets $(G,s,t,f)$
  as input, where $f$ is some flow in $G$ from $s$ to~$t$, and will
  output $(G,s,t,f')$, where $f'$ is a flow of value $|f|+1$ (provided
  such a flow exists, otherwise $f' = f$). If there is no path from $s$
  to $t$ in the residual network~$R_f$ (an $\Class{NL}$-machine 
  can easily check this), set $f' = f$. Otherwise, use
  Lemma~\ref{lemma:path-nl} to compute such a path (called an
  \emph{augmenting path}) and output the flow~$f'$ that corresponds
  to~$f$ augmented by the path (initially set $f'=f$ and then for
  every edge $(u,v)$ on
  the path set $f'(u,v)=1$ if $(u,v)\in E$; and $f'(v,u)=0$ otherwise).
  
  By setting $r(k)=k$, we get a
  value-$k$ or maximum-value flow from $s$ to $t$. 
\end{proof (possibly in appendix)}

Let $\PLang[k]{flow} = \bigl\{
\bigl((G,s,t),k\bigr) \mid{}$there is a 0-1-flow~$f$ from $s$ to
$t$ in $G$ with $|f| \le k\bigr\}$ be the corresponding parameterized decision
problem.

\begin{corollary}\label{corollary:0-1-flow}  
  $\PLang[k]{flow}\in \Para\Class{NL}^\uparrow.$
\end{corollary}

The following corollary observes that instead of starting with the
empty flow we can also 
start with an arbitrary flow~$f$ and augment it $k$~times:

\begin{corollary}\label{corollary:flow-update}
  There is a parameterized function in $\Para\Class{FNL}^\uparrow$
  that maps $((G,s,t,f),k)$, where $f$ is an $s$-$t$-flow in~$G$, to
  an $s$-$t$-flow $f'$ in~$G$ of value $|f|+k$, if it exists, or to a
  maximum flow otherwise.  
\end{corollary}

\begin{remark}
  While Theorem~\ref{theorem:flow} and the corollaries~\ref{corollary:0-1-flow}
  and~\ref{corollary:flow-update} only speak about 0-1-flows, it is easy
  to see that the same techniques can be used to compute flows in
  networks with \emph{fixed constant capacities:} just replace each
  edge with capacity~$c$ by $c$ parallel edges and divide each of
  these edges with a fresh vertex afterwards. In particular,
  Corollary~\ref{corollary:flow-update} can also be used to augment
  \emph{half-integral flows} in networks with fixed maximum capacity.
\end{remark}

\subsubsection{Linear Programs for Vertex Cover and Matching}
To prove Theorem~\ref{theorem:vc-above-matching}, we will study a more
general problem and obtain the theorem as a simple corollary: Instead
of using matchings as proofs for lower bounds for the vertex cover
problem, we use fractional solutions of LP-relaxations. Let us fix some
notations: For a linear program $\Pi$ let $\vars(\Pi)$ be the set of variables
occurring in $\Pi$. A \emph{solution} for $\Pi$ is an assignment
$\alpha\colon\vars(\Pi)\rightarrow\mathbb{Q}$ that satisfies
all inequalities, and the \emph{solution value} (or just \emph{value})
$|\alpha|$ of $\alpha$ is the value of the optimization function
under~$\alpha$. An \emph{optimal solution} is an 
assignment that has the minimum (or maximum) solution value over all
possible assignments. We say an assignment $\alpha$ is \emph{integral} 
if $\alpha\colon\vars(\Pi)\rightarrow\mathbb{N}$ for all
$x\in\vars(\Pi)$; $\alpha$ is \emph{half-integral} if all
$\alpha(x)$ are half-integral, meaning $\alpha(x) = i/2$ for some
$i\in\mathbb N$; otherwise $\alpha$ is \emph{fractional}. Let $\operatorname{opt}_{\mathbb Q}(\Pi)$,
$\operatorname{opt}_{\mathbb N/2}(\Pi)$, and
$\operatorname{opt}_{\mathbb N}(\Pi)$ denote the optimal value of a
fractional, half-integral, and integral solution for~$\Pi$, respectively. 
We are interested in the following two linear programs:
\begin{definition}[Linear Program $\LP{VC}(G)$ for Vertex Cover of a
    Graph $G = (V,E)$]
  \hfil\par
  \begin{tabular}{lll}
    Minimize            $\sum_{v\in V} x_v$                         
    subject to         & $x_u+x_v\geq 1$    & for all $\{u,v\}\in E$, \\
                       & $0\leq x_v\leq 1$  & for all $v\in V$.
  \end{tabular}
\end{definition}
\begin{definition}[Linear Program $\LP{M}(G)$ for Matching of a Graph $G = (V,E)$]
  \hfil\par
  \begin{tabular}{lll}
    Maximize           $\sum_{e\in E} y_e$                        
    subject to        & $\sum_{v\in e}y_e\leq 1$ & for all $v\in V$, \\
                      & $0\leq y_e\leq 1$        & for all $e\in E$.
  \end{tabular}
\end{definition}
A vertex cover of~$G$ naturally corresponds to an
integral solution $\alpha_{\mathbb N}$ of $\LP{VC}(G)$ and a matching
corresponds to an integral solution $\beta_{\mathbb N}$ of
$\LP{M}(G)$ (the index ``$\mathbb N$'' emphasizes that the
solution is integral). In particular, $\operatorname{opt}_{\mathbb
  N}(\LP{VC}(G))$ and $\operatorname{opt}_{\mathbb N}(\LP{M}(G))$ are
the sizes of a minimum vertex cover and a maximum matching of~$G$,
respectively. The programs are dual to each other, which implies that
their optimal fractional solutions have the same value.
\begin{fact}[Nemhauser-Trotter Theorem \cite{NemhauserT75},
    {\cite[Chapter~2]{CyganFKLMPPS15}}]\label{fact:nemhauser}
  Let $G=(V,E)$ be a graph. Then $\LP{VC}(G)$ and $\LP{M}(G)$
  have solutions $\alpha$ and $\beta$, respectively, with the
  following properties:
  \begin{enumerate}
    \item $\operatorname{opt}_{\mathbb Q}(\LP{VC}(G))=|\alpha| = |\beta| = \operatorname{opt}_{\mathbb Q}(\LP{M}(G))$, 
    \item $\alpha$ and $\beta$ are \emph{half-integral},
    \item there is an optimal \emph{integral} solution~$\gamma$ for
      $\LP{VC}(G)$ such that for $v\in V$ with
      $\alpha(x_v)\neq 1/2$ we have $\gamma(x_v)=\alpha(x_v)$ (that
      is, $\gamma$ equals $\alpha$ on its integral part). 
  \end{enumerate}
\end{fact}

Fact~\ref{fact:nemhauser} implies that the following (in)equalities hold, where
$\alpha_{\mathbb N}$ and $\alpha_{\mathbb N/2}$ are arbitrary integral
and half-integral solutions for $\LP{VC}(G)$ and $\beta_{\mathbb N}$
and $\beta_{\mathbb N/2}$ correspondingly for $\LP{M}(G)$:

\begin{align}
  \begin{tikzpicture}[
      every node/.style={anchor=base,inner sep=0pt,minimum height=2mm},
      x=2.9cm,
      y=4mm,
      baseline=4mm,
    ]
    \path (4,2.75);
    \node (b1) at (0.45,2) {\smash{$|\beta_{\mathbb N}|$}};
    \node (b2) at (1.25,0) {\smash{$|\beta_{\mathbb N/2}|$}};
    \node (m1) at (1.1,2) {\smash{$\operatorname{opt}_{\mathbb N}(\LP{M}(G))$}};
    \node (m2) at (2,1) {\smash{$\overbrace{\operatorname{opt}_{\mathbb N/2}(\LP{M}(G))}^{(*)}$}};
    \node (v1) at (3,1) {\smash{$\operatorname{opt}_{\mathbb N/2}(\LP{VC}(G))$}};
    \node (v2) at (4,2) {\smash{$\overbrace{\operatorname{opt}_{\mathbb N}(\LP{VC}(G))}^{(**)}$}};
    \node (a1) at (4.65,2) {\smash{$|\alpha_{\mathbb N}|$}};
    \node (a2) at (3.75,0) {\smash{$|\alpha_{\mathbb N/2}|$}};
    \foreach \a/\b/\what in {b1/m1/\le,
      m1/m2/\le,m2/v1/=,v1/v2/\le,v2/a1/\le,
      b2/m2/\le,v1/a2/\le}
    { \path (\a.east) -- node[sloped] {\smash{$\what$}} (\b.west);}    
  \end{tikzpicture}
  \label{eq:lps}
\end{align}

The parameter of
$\PLang[g]{vc-above-matching}$ is the difference 
between the upper left value~$|\beta_{\mathbb N}|$, which is the size
of some matching of~$G$, and $(**)$, which is the size of a minimum
vertex cover of~$G$. When working with linear programs, it is natural
to work with a different (``better'') parameter, namely the difference between the lower left
value~$|\beta_{\mathbb N/2}|$ and~$(**)$:

\begin{problem}[{$\PLang[g]{vc-above-relaxed-matching}$}]\label{problem:vc-lp}
\begin{parameterizedproblem}
  \instance A graph $G=(V,E)$, a half-integral solution
  $\beta_{\mathbb N/2}$ for
  $\LP{M}(G)$, and a number $g$.
  \parameter $g$
  \question Is there a set $S\subseteq V$ with $|S|\leq
  |\beta_{\mathbb N/2}|+g$ and $e\cap
  S\neq\emptyset$ for every $e\in E$? 
\end{parameterizedproblem}
\end{problem}

\subsubsection{An FPT-Algorithm for Solving VC Above Half-Integral Matching}
\label{section:ftp-algo}

Let us briefly review how one usually shows
$\PLang[g]{vc-above-relaxed-matching} \in \Para\Class P$:
\begin{description}
\item[Step 0: Computing an Optimal Half-Integral Solution.] Compute an optimal half-integral
  solution $\alpha$ for $\LP{VC}(G)$ in polynomial-time ($|\alpha|$ has the value $(*)$ in~\eqref{eq:lps}).
\item[Step 1: Reduction to the All-1/2-Solution.] We
  turn $\alpha$ into an ``all-$1/2$-solu\-tion'' $\alpha \equiv 1/2$,
  meaning  $\alpha(x_v) = 1/2$ for all $v\in V$. Fact~\ref{fact:nemhauser} tells us that vertices with $\alpha(x_v) = 0$ are not part of an optimal vertex
  cover while vertices with $\alpha(x_v) = 1$ are. Thus, we
  can delete all these vertices and continue with the same parameter~$g$ (the integrality
  excess does not change). Note that $\alpha$
  restricted to the new graph (which we still call $G$) is constantly $1/2$.  
\item[Step 2: Making the All-1/2-Solution Unique.] Now
  $\alpha \equiv 1/2$ is an optimal solution, but there may be other
  optimal half-integral solutions. (For instance, the all-1/2-solution
  is an optimal solution for any even cycle, but so is the integral
  solution $\alpha(i) = (i \bmod 2)$.) We can check in polynomial time
  whether $\alpha$ is the unique optimal solution as follows: Test for
  every $x_v$ whether
  $\operatorname{opt}_{\mathbb Q}(\LP{VC}(G)) =
  \operatorname{opt}_{\mathbb Q}(\LP{VC}(G - \{v\})) + 1$.  If so,
  there is an optimal solution other than $\alpha$~that assigns $1$ to
  $x_v$. We remove $v$ from $G$ using Fact~\ref{fact:nemhauser}, leave
  $g$ untouched, and repeat until the all-1/2-solution is the only
  optimal solution.
\item[Step 3: Branching.] Suppose we knew that some
  vertex $v \in V$ is part of an optimal vertex cover of~$G$. Then
  $\operatorname{opt}_{\mathbb N}(G - \{v\}) = 
  \operatorname{opt}_{\mathbb N}(G) - 1$ while
  $\operatorname{opt}_{\mathbb Q}(G - \{v\}) =
  \operatorname{opt}_{\mathbb Q}(G) - 1/2$. This means that the
  integrality excess of $G - \{v\}$ is reduced by $1/2$ compared
  to~$G$. Of course, we do not know which vertices are part
  of an optimal vertex cover, but we can find them using
  \emph{branching:} Pick an arbitrary edge $\{u,v\} \in E$ and
  recursively run the whole algorithm (starting from Step~1 once
  more) for $G - \{u\}$ and $G - 
  \{v\}$, but now for the parameter $g-1/2$ (the parameter
  should actually be an integer, but it is convenient for the
  recursion to allow integers divided by~$2$ as parameters in
  this setting).
\end{description}
It is now easy to see that the depth of the search tree of the above
algorithm is $2g$, so the total runtime is $4^g
\cdot n^{O(1)}$.

\subsubsection{A Parallel Algorithm for VC Above Half-Integral Matching}

In this section we parallelize the different steps
sketched above for solving 
Problem~\ref{problem:vc-lp}. This yields the following theorem, of which Theorem~\ref{theorem:vc-above-matching} is a
corollary\iftcsautomove~--~the formal proof can be
found in the appendix, but the content of this section should be read before\fi:

\begin{theorem (proof possibly in appendix)}\label{theorem:vc-above-hi-matching}
  $\PLang[g]{vc-above-relaxed-matching} \in \Para\Class{NL}^\uparrow$.
\end{theorem (proof possibly in appendix)}

While steps 1 and~3 are easy to parallelize (search trees can
be traversed in parallel), steps~0 
and~2 are not. They either involve open 
problems (like computing optimal solutions for $\LP{VC}(G)$ in
parallel) or are very sequential (like the iterative removal of
vertices in step~2).

\subparagraph*{Parallelizing Step~0: Computing an Optimal
  Half-Integral Solution.}
Given a half-integral solution $\beta$ of $\LP{M}(G)$, we wish
to compute an optimal half-integral solution $\alpha$ of
$\LP{VC}(G)$. A $\Para\Class P$-machine could just ignore $\beta$ and solve the linear program, but
we only have a $\Para\Class{NL}^\uparrow$-machine.
The core idea we use was developed by Iwata, Oka, and
Yoshida~\cite{IwataOY14} in the context of a linear-time algorithm:
One can encode an (optimal) solution of $\LP{M}(G)$ into a (maximum)
flow in the so-called \emph{Hochbaum network.} More
crucially, we can obtain an (optimal) solution for $\LP{M}(G)$
\emph{and} $\LP{VC}(G)$ from a (maximum) flow in this network. 

In detail, for a graph $G=(V,E)$ the
\emph{Hochbaum network} is the digraph $H = (V', E')$ with~$V'$
consisting of $V_1 = \{\,v_1 \mid v \in V\,\}$ 
and $V_2 = \{\,v_2 \mid v \in V\,\}$ plus the two vertices
$s$ and~$t$. The edge set is 
\(
E' = \{\, (s,v_1)   \mid  v\in V       \,\}
\cup \{\, (u_1,v_2) \mid  \{u,v\}\in E \,\}
\cup \{\, (v_2,t)   \mid  v\in V       \,\}
\),
i.\,e., from $s$ we get to all vertices in $V_1$, then we can
cross from $u_1$ to $v_2$ exactly if $\{u,v\} \in E$ (and then also
from $v_1$ to $u_2$), and from all vertices in $V_2$ we can get to~$t$. 

\begin{fact}[\cite{Hochbaum02,IwataOY14}]\label{fact:betaflow} 
  Let $G=(V,E)$ be a graph and $H=(V',E')$ be its
  Hochbaum network. 
  \begin{enumerate}
  \item If $\beta$ is a solution of $\LP{M}(G)$, then the mapping
    $f_\beta(s,v_1)=f_\beta(v_2,t)=\sum_{w\in N(v)}\beta(y_{\{v,w\}})$ and
    $f_\beta(u_1,v_2)=f_\beta(v_1,u_2) =\beta(y_{\{u,v\}})$ is an $s$-$t$-flow with
    $|f_\beta|=2|\beta|$ in~$H$. 
  \item If $f$ is an $s$-$t$-flow in~$H$, then
    $\beta_f(y_{\{u,v\}})=\frac{1}{2}\big(f(u_1,v_2)+f(v_1,u_2)\big)$ is a
    solution for $\LP{M}(G)$ with $|\beta_f|=|f|/2$.
  \end{enumerate}
\end{fact}

Note that, in particular, $\beta$ is an optimal solution of
$\LP{M}(G)$ iff $f_\beta$ is maximal and, \emph{vice versa,} $f$ is a maximal
flow iff $\beta_f$ is an optimal
solution. Figure~\ref{figure:graph-Hochbaum-triple} \iftcsautomove in
the appendix\fi\ illustrates
these definitions and the interplay between solutions for $\LP{M}(G)$
and flows in the corresponding Hochbaum network. Since the translation
between flows and solutions is computationally easy, we freely switch
between flows and solutions for $\LP{M}(G)$ as needed. 

\begin{figure (possibly in appendix)}[htbp]
  \begin{tikzpicture}[
    vertex/.style = {
      draw,
      circle,
      inner sep     = 0pt,
      minimum width = 5pt,
      semithick
    },
    every label/.style = {
      color = ba.orange,
      font  = \small,
      label position = 135,
      label distance = -1mm
    }
    ]
    
    \draw[rectangle, rounded corners, color=lightgray] (0,0) rectangle (6,5);
    \draw[rectangle, rounded corners, color=lightgray] (8,0) rectangle (14,5);

    \node[anchor=west] at (0, 5.25) {Graph $G=(V,E)$};
    \node[anchor=west, baseline] at (8, 5.25) {Hochbaum network $H=(V',E')$};
    
    \node at (3,2.5) {\tikz{
        \node[vertex, label=$a$]     (a) at (2, 3) {};
        \node[vertex, label=$b$]     (b) at (2, 1) {};
        \node[vertex, label={[label distance=0mm]90:$c$}]  (c) at (3, 2) {};
        \node[vertex, label=$d$]     (d) at (4, 3) {};
        \node[vertex, label=225:$e$] (e) at (4, 1) {};
        \weightededge[color=ba.red]{a}{b}{\sfrac{1}{2}}
        \weightededge{a}{c}{\sfrac{1}{2}}
        \weightededge{b}{c}{\sfrac{1}{2}}
        \weightededge[color=ba.red]{c}{d}{0}
        \weightededge{c}{e}{0}
        \weightededge{d}{e}{1}
      }};

    \node at (11,2.5) {\tikz{
        \node[vertex, label=$a_1$] (a1) at (1,4) {};
        \node[vertex, label=$b_1$] (b1) at (1,3) {};
        \node[vertex, label=$c_1$] (c1) at (1,2) {};
        \node[vertex, label={[label distance=0mm]90:$d_1$}] (d1) at (1,1) {};
        \node[vertex, label={[label distance=0mm]180:$e_1$}] (e1) at (1,0) {};
        \node[vertex, label=45:$a_2$] (a2) at (3,4) {};
        \node[vertex, label=45:$b_2$] (b2) at (3,3) {};
        \node[vertex, label=45:$c_2$] (c2) at (3,2) {};
        \node[vertex, label={[label distance=0mm]90:$d_2$}] (d2) at (3,1) {};
        \node[vertex, label={[label distance=0mm]0:$e_2$}] (e2) at (3,0) {};
        \node[vertex, label=$s$]    (s) at (-0.5,2) {};
        \node[vertex, label=45:$t$] (t) at ( 4.5,2) {};
        
        \arc[color=ba.red]{a1}{b2}{\sfrac{1}{2}}
        \arc{a1}{c2}{\sfrac{1}{2}}
        
        \arcclose[color=ba.red, bend right=10]{b1}{a2}{\sfrac{1}{2}}
        \arc{b1}{c2}{\sfrac{1}{2}}

        \arcclose[bend right=10]{c1}{a2}{\sfrac{1}{2}}
        \arc[bend right=25]{c1}{b2}{\sfrac{1}{2}}
        \draw[color=ba.red, semithick, ->, >={[round,sep]Stealth}] (c1) to (d2);
        \draw[semithick, ->, >={[round,sep]Stealth}] (c1) to (e2);
        
        \draw[color=ba.red, semithick, ->, >={[round,sep]Stealth}] (d1) to (c2);
        \arc{d1}{e2}{1}
        
        \draw[semithick, ->, >={[round,sep]Stealth}] (e1) to (c2);                
        \arc{e1}{d2}{1}

        \arc[color=ba.red]{s}{a1}{1}
        \arc[color=ba.red]{s}{b1}{1}
        \arc[color=ba.red]{s}{c1}{1}
        \arc[color=ba.red]{s}{d1}{1}
        \arc{s}{e1}{1}

        \arc[color=ba.red,<-]{t}{a2}{1}
        \arc[color=ba.red,<-]{t}{b2}{1}
        \arc[color=ba.red,<-]{t}{c2}{1}
        \arc[color=ba.red,<-]{t}{d2}{1}
        \arc[<-]{t}{e2}{1}        
      }};
  \end{tikzpicture}
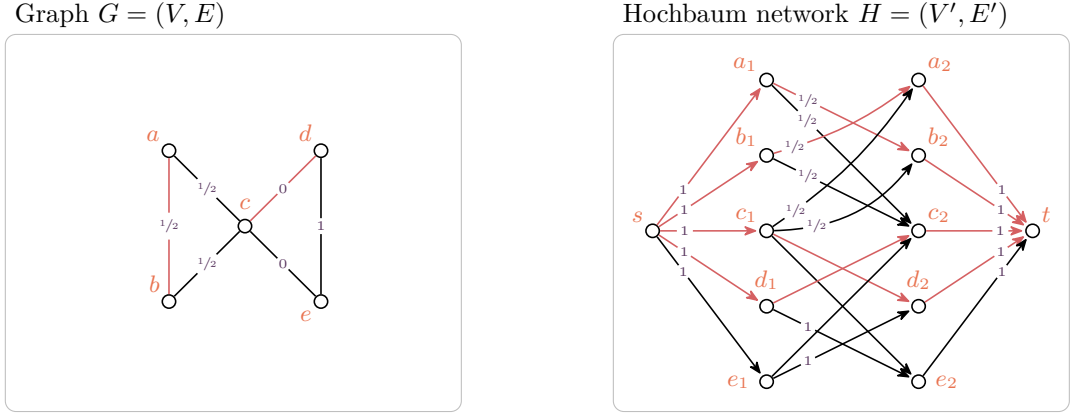
  \caption{The left side shows a graph $G=(V,E)$ on five vertices
    $V=\{a,b,c,d,e\}$. On the edges a \emph{half-integral} solution
    $\beta$ of
    $\LP{M}(G)$ of value $\operatorname{opt}_{\mathbb
      N/2}(\LP{M}(G))=2.5$ is illustrated. The two red edges
    constitute an optimal \emph{integral} solution for $\LP{M}(G)$.
    The right side shows the Hochbaum network $H=(V',E')$
    corresponding to $G$. The edges are labeled with a maximum flow $f_{\beta}$ of value
    $|f_{\beta}|=5$ that corresponds to~$\beta$. The integral solution
    (the red maximum matching)
    corresponds to the flow of value four that sends one unit over
    every red edge (which is not maximal).
  }  
  \label{figure:graph-Hochbaum-triple}
\end{figure (possibly in appendix)}

\begin{lemma (proof possibly in appendix)}\label{lemma:half}
  There is a function in $\Para\Class{FNL}^\uparrow$ that maps
  $((G,\beta),g)$, consisting of a graph~$G$, a half-integral
  solution $\beta$ of $\LP{M}(G)$, and a number~$g$, to
  an \emph{optimal} half-integral solution $\beta'$ of~$G$, provided
  such a solution with $|\beta'| \le |\beta| + g$ exists.
\end{lemma (proof possibly in appendix)}

\begin{proof (possibly in appendix)}[Proof of Lemma~\ref{lemma:half}]
  The mapping of $(G, \beta)$ to the Hochbaum network~$H$ and the
  flow~$f_\beta$ from Fact~\ref{fact:betaflow} can easily be done in (even
  deterministic) logarithmic space. The flow~$f_\beta$ does not have
  to be a maximal flow, but we can turn it into a maximal flow using
  Corollary~\ref{corollary:flow-update}: This corollary states that a
  $\Para\Class{FNL}^\uparrow$-machine can map $f_\beta$ to a flow~$f'$
  of value $|f_\beta| + 2g + 1$, if such a flow exists, or to a
  maximum flow~$f'$ otherwise.

  If a flow of value $|f_\beta| + 2g + 1$ exists,
  Fact~\ref{fact:lpToflow} tells us that $\beta_{f'}$ is a 
  solution of $\LP{M}(G)$ of value $|\beta_{f'}| = |f'|/2 =
  |f_\beta|/2 + g + 1/2 = |\beta| + g + 1/2$. In particular,
  $|\beta| + g < |\beta_{f'}| \le \operatorname{opt}_{\mathbb
    Q}(\LP{M}(G))$ and, thus, no optimal solution 
  with $|\beta| + g \ge |\beta'|$ exists. Hence, we 
  output the error symbol.

  If there is no flow of value $|f_\beta| + 2g + 1$, we know that $f'$
  is a maximum $s$-$t$-flow in $H$ and, by Fact~\ref{fact:betaflow},
  we can output $\beta_{f'}$ as optimal solution for $\LP{M}(G)$.
\end{proof (possibly in appendix)}

Lemma~\ref{lemma:half} provides a reduction rule for
$\PLang[g]{vc-above-relaxed-matching}$: We can map
$((G,\beta),g)$ to $((G,\beta'),g-|\beta'|+|\beta|)$ such that $\beta'$ is optimal.
Since we will often use triples $(G,H,f)$
where $G = (V,E)$ is a graph, $H = (V',E')$ is its
Hochbaum network, and $f$ is a maximum flow in~$H$, let us call such a
triple a \emph{graph-Hochbaum-flow triple.} 

We now have a way of computing an optimal solution $\beta'$ for
$\LP{M}(G)$, but for the next steps of our algorithm, we need a
half-integral solution $\alpha$ of $\LP{VC}(G)$. Fortunately, there is
another observation that shows how a maximum flow $f$ can be used to
derive an optimal solution $\alpha_f$ for $\LP{VC}(G)$ (note that
this solution is trivially half-integral): 

\begin{fact}[\cite{Hochbaum02,IwataOY14}]\label{fact:lpToflow} 
  Let $(G,H,f)$ be a graph-Hochbaum-flow triple. Let $X\subseteq V'$
  be the set of vertices reachable from $s$ in the residual network
  $R_f$. Then 
  \begin{align*}
    \alpha_f(x_v) =
    \begin{cases}
      0           & \text{if $v_1\in X$ and $v_2\not\in X$,} \\
      1           & \text{if $v_1\not\in X$ and $v_2\in X$,} \\
      1/2  & \text{otherwise,}                             \\
    \end{cases}  
  \end{align*}
  is an optimal solution for $\LP{VC}(G)$.
\end{fact}

\begin{lemma (proof possibly in appendix)}\label{lemma:beta-to-alpha}
  There is a function in $\Class{FNL}$ that maps
  $(G,\beta)$, consisting of a graph and an \emph{optimal}
  half-integral solution of $\LP{M}(G)$, to
  an optimal half-integral solution $\alpha$ of $\LP{VC}(G)$.
\end{lemma (proof possibly in appendix)}

\begin{proof (possibly in appendix)}[Proof of Lemma~\ref{lemma:beta-to-alpha}]
  Use Fact~\ref{fact:betaflow} to obtain an optimal flow $f_\beta$ in
  the Hochbaum network from $\beta$ and use Fact~\ref{fact:lpToflow}
  to obtain $\alpha_{f_\beta}$ from this flow.
\end{proof (possibly in appendix)}

Together, Lemmas \ref{lemma:half} and~\ref{lemma:beta-to-alpha}
clearly allow us to perform Step~0 of the computation (namely the
computation of an optimal solution $\alpha$ of $\LP{VC}(G)$) using a
$\Para\Class{FNL}^\uparrow$-machine.

\subparagraph*{Parallelizing Step~1: Reduction to the
  All-1/2-Solution.} 

The next step turns the half-integral
solution~$\alpha$ into an all-1/2-solution by deleting all
vertices $v$ for which $\alpha(x_v) \neq 1/2$. Clearly,
this can be done in parallel.
Note that here we really need an optimal
solution $\alpha$ of $\LP{VC}(G)$ rather than a solution $\beta$ of
$\LP{M}(G)$: Only $\alpha$ tells us which vertices can be removed.

\subparagraph*{Parallelizing Step~2: Making the All-1/2-Solution
  Unique.} 

The sequential method described in Section~\ref{section:ftp-algo} for
implementing Step~2 is exactly that: highly \emph{sequential.} It is
not difficult to construct a graph for which the number of iterations
used by this method is linear in the graph size -- just consider a large
matching: The all-1/2-solution is an optimal solution, but in each
iteration of Step~2 only one edge will be removed from the graph. Even worse, after the
removal of a vertex it might be necessary to recompute the optimal
solution~$\alpha$.

For a parallel algorithm, we need some further insights from the work
of Iwata, Oka, and Yoshida~\cite{IwataOY14}. Let us start with some
definitions, which adapt their ideas to our context:

\begin{definition}\label{definition:removable}
  Let $(G,H,f)$ be a graph-Hochbaum-flow triple. A set $S \subseteq
  V'$ is \emph{loose} if the following holds:
  \begin{enumerate}
  \item $S$ is a strongly connected component of the residual
    graph $R_f = (V',E'_f)$.
  \item $\{\,v \in V \mid v_1 \in S\,\}$ and $\{\,v \in V \mid v_2 \in S\,\}$
    are disjoint.
  \end{enumerate}
  We call a loose set \emph{removable,} if the following holds
  additionally: 
  \begin{enumerate}\setcounter{enumi}{2}
  \item There are no edges leaving $S$ in $R_f$, i.\,e., no edges
    $(x,y) \in E'_f$ with $x \in S$ and $y \notin S$.
  \end{enumerate}
\end{definition}

\begin{definition}\label{definition:remove}
  Let $(G,H,f)$ be a graph-Hochbaum-flow triple and $S \subseteq V'$
  be a removable set. \emph{Removing~$S$} yields the following triple
  $(G^{-S}, H^{-S}, f^{-S})$:
  \begin{enumerate}
  \item $G^{-S} = G - \{\,v\in V\mid v_1\in S\vee \exists w\in N(v) (w_1\in S)\,\}$,
  \item $H^{-S} = H - S$,
  \item $f^{-S}$ is the flow induced on the vertices of $H^{-S}$.
  \end{enumerate}
\end{definition}

Intuitively, $(G^{-S}, H^{-S}, f^{-S})$ should also be a
graph-Hochbaum-flow triple and this is the case, at least if $\alpha
\equiv 1/2$ is an optimal solution:

\begin{fact}[{\cite[Corollary~4.2 and the subsequent discussion]{IwataOY14}}]\label{fact:iwata}
  Let $(G,H,f)$ be a graph-Hochbaum-flow triple such that $\alpha
  \equiv 1/2$ is an optimal solution of $\LP{VC}(G)$.
  \begin{enumerate}
  \item If there is no removable set~$S$, then $\alpha \equiv 1/2$ is
    the only optimal solution for $\LP{VC}(G)$.
  \item If there is a removable set~$S$, then $(G^{-S}, H^{-S},
    f^{-S})$ is a graph-Hochbaum-flow triple and $G^{-S}$ has the same
    integrality excess as~$G$.
  \end{enumerate}
\end{fact}

While the fact tells us which vertices we should remove from~$G$, it
does not tell us which will be part of the vertex cover. This can
easily be fixed, however: When $S$ is removed, we can set
$\beta(x_v)=0$ for all $v_1\in S$ and $\beta(x_v)=1$ for all
$v\in V$ for which there is a $w\in N(v)$ with
$w_1\in S$, see the discussion after Lemma~4.6 in~\cite{IwataOY14} for details.

Using Fact~\ref{fact:iwata}, an $\Class{NL}$-machine can test whether $\alpha \equiv 1/2$ is
the only optimal solution of $\LP{VC}(G)$ by looking for a
removable~$S$. Furthermore, the machine can iteratively remove such
sets until the all-$1/2$-solution is the only optimal half-integral
solution. This may seem similarly sequential as the
repetitive removal of vertices in Step~2, but it turns out that we can
remove everything in a single run:

\begin{lemma (proof possibly in appendix)}\label{lemma:all-iwata}
  There is a function in $\Class{FNL}$ that gets a graph-Hochbaum-flow
  triple $(G,H,f)$ as input and outputs the graph-Hochbaum-flow triple
  $(G^{-}, H^{-}, f^{-})$ resulting from iteratively removing
  removable sets as long as they exist.
\end{lemma (proof possibly in appendix)}

\begin{proof (possibly in appendix)}[Proof of Lemma~\ref{lemma:all-iwata}]
  Consider the acyclic digraph $D$ of all strongly connected
  components~$C$ of~$H$. Some of these components will be loose sets
  (see Definition~\ref{definition:removable}) and if they are also sinks in
  $D$, they are one of the (initial) removable sets of~$H$. Note that
  removing one of these loose sinks does not change the fact that the
  other loose sinks are (still) removable sets in the resulting
  graph-Hochbaum-flow triple. Removing loose sinks from~$H$ and~$D$ may produce new loose sinks, but 
  these sets were already loose sets in the original~$H$
  (Definition~\ref{definition:removable} is ``local'' in the
  sense that only properties of vertices within the strongly
  connected component are relevant).

  This leads to a rule for determining the set $Q$ of all
  vertices that will (eventually) be removed as part of the iterative
  removal of removable sets: $Q$ contains all vertices that are an element of a
  loose set from which only loose sets are reachable in~$D$. This test can be implemented by an
  $\Class{FNL}$-machine and the claim follows with $(G^{-Q}, H^{-Q}, f^{-Q})$.
\end{proof (possibly in appendix)}

\subparagraph*{Parallelizing Step~3: Branching.}

As mentioned earlier, the branching step is easy to parallelize, as
the two children in the search tree can be explored in parallel.
Branching also fits nicely into our framework of
the up-class $\Para\Class{FNL}^\uparrow$, which arises from
parameter-dependent-many iterations of a linear function in $\Para\Class{FNL}$:
In each iteration a list of instances is on the input tape and this
list is
mapped to at most twice as many new instances on the output tape, but
with a reduction of the parameter in all these instances.

\begin{proof (possibly in appendix)}[Proof of Theorem~\ref{theorem:vc-above-hi-matching}]
{ Let $((G,\beta),g)$ be given as
  input, where $G = (V,E)$ is an undirected graph, $\beta$ is a
  half-integral solution of $\LP{M}(G)$, and $g$ is a parameter.

  To show that a problem is in $\Para\Class{NL}^\uparrow$, we must
  specify an initial function and an iteration function, both in
  $\Para\Class{FNL}$. In our case the initial function simply maps
  $((G,\beta),g)$ to the single-element list
  $\bigl(((G,\beta),g)\bigr)$. This list, which will change after each
  application of the iteration function, will satisfy the following
  invariant: The 
  original instance $((G,\beta),g)$ is a positive instance iff at
  least one instance in the list is a positive instance. Clearly,
  after the application of the initial function, this invariant is
  true. 

  The iteration function gets a list
  $\bigl(((G_1,\beta_1),g_1),\dots,((G_l,\beta_l),g_l)\bigr)$ as input
  and will output a new list of such pairs that is at most 
  twice as long (which will ensure that the iteration function is
  linear, see Definition~\ref{definition:up}). When processing the pairs, the
  iteration function may 
  notice that one of the pairs is a positive instance. Because of the
  invariant, the iteration function can now immediately output
  ``yes'' (formally, it outputs $(1,0)$ and further iterations do
  nothing except for copying
  this tuple to their output tape). It may also happen that the
  list becomes empty (at the latest after $2g+1$
  iterations), in which case the invariant implies that the original
  instance was a negative instance and the iterations function
  immediately outputs ``no'' in the form of $(0,0)$ (and once more
  further iterations do not modify this).

  We now describe how the iteration function processes a pair
  $((G_i,\beta_i),g_i)$ in the list, i.\,e., which new pairs are
  added to the output list (if any). For Definition~\ref{definition:up}, we
  have to implement the iterator function in $\Para\Class{FNL}$, but
  Lemma~\ref{lemma:closure-up} allows us to use
  $\Para\Class{FNL}^\uparrow$-transformation instead, as long as the
  initial functions are linear (which they are).
  \begin{description}
  \item[Step 0, first part.] Apply Lemma~\ref{lemma:half} to $((G_i,\beta_i),g_i)$. This
    will yield a new instance $((G_i,\beta'),g')$ such that $\beta'$
    is an optimal solution of $\LP{M}(G_i)$ -- or an error symbol, in
    which case we know that $((G_i,\beta_i),g_i)$ was a no-instance
    and we can skip it.
  \item[Step 0, second part.] Apply Lemma~\ref{lemma:beta-to-alpha} to obtain an optimal
    solution $\alpha$ for $\LP{VC}(G_i)$.
  \item[Step 1.] Remove all vertices $v \in V$ from $G_i$ with $\alpha(v) \neq
    1/2$, yielding the graph~$G'$.
  \item[Step 2.] Compute the graph-Hochbaum-flow triple $(G',H',f')$
    and apply Lemma~\ref{lemma:all-iwata} to it. This yields a
    graph-Hochbaum-flow triple $(G^-,H^-,f^-)$ such that (i) the
    integrality excess of $G^-$ is the same as that of $G$ and, hence,
    $((G^-,\beta_{f^-}),g')$ is an element of
    $\PLang[g]{vc-above-relaxed-matching}$ iff $((G',\beta_{f'}),g')$
    is, and (ii) $\alpha \equiv 1/2$ is the only optimal half-integral
    solution of $\LP{VC}(G^-)$.
  \item[Step 3.] If in $((G^-,\beta_{f^-}),g')$ the graph $G^-$ contains
    no edges and $g' \ge 0$, we have found a yes-instance and can
    stop. Likewise, if $g'< 0$, we have a no-instance and can also
    stop. Otherwise, we branch by picking an arbitrary edge $e = \{u,v\}$ 
    in~$G^-$ and, starting with~$u$, consider the graph $G_u = G^-
    -  \{u\}$. In the corresponding Hochbaum network $H_u$ the
    vertices $u_1$ and $u_2$ will be missing. Consider the flow $f_u$ that
    is obtained from $f^-$ by removing any flow through $u_1$
    or~$u_2$. Then $|f_u| \ge |f^-| - 2$ and $|f_u|$ can be at
    most $2$ below the value of a maximum flow in
    $H_u$. Lemma~\ref{lemma:half} allows us to restore the maximality by computing a
    maximum flow $f_u'$ in $H_u$. We add $((G_u,\beta_{f_u'}),g'-1/2)$
    to the list. Then we repeat the whole process with $v$ and also
    add $((G_v,\beta_{f_v'}),g'-1/2)$ to the list.

    To see that the branching is correct and upholds the invariant, 
    suppose $((G^-,\beta_{f^-}),g')$ is a yes-instance, i.\,e., the
    integrality excess of $G^-$ is at most $g'$. For the edge $\{u,v\}$
    one of the vertices must be in a minimal vertex cover -- suppose
    it is~$u$. Then
    \begin{align*}
      \operatorname{opt}_{\mathbb N}(\LP{VC}(G^-)) &=
      \operatorname{opt}_{\mathbb N}(\LP{VC}(G^--\{u\})) + 1, \\
      \operatorname{opt}_{\mathbb Q}(\LP{VC}(G^-)) &=
      \operatorname{opt}_{\mathbb Q}(\LP{VC}(G^--\{u\})) + 1/2.
    \end{align*}
    To see the last equality, observe that if we had
    $\operatorname{opt}_{\mathbb Q}(\LP{VC}(G^-)) =
    \operatorname{opt}_{\mathbb Q}(\LP{VC}(G^--\{u\})) + 1$, then 
    any optimal solution $\alpha$ for $\LP{VC}(G^--\{u\})$ could be
    augmented to an optimal solution for $\LP{VC}(G^-)$ by setting
    $\alpha(u) = 1$, contradicting the assumption that $\alpha \equiv
    1/2$ is the only optimal solution of $\LP{VC}(G^-)$. The two
    equalities taken together show that the integrality excess of
    $\LP{VC}(G^--\{u\})$ is, indeed, $1/2$ less than that of~$G^-$.
    \qedhere
  \end{description}}
\end{proof (possibly in appendix)}

\subsection{Dual Parameterization When Every Variables Occur at Most Twice}
\label{section:dual-twice}

A formula $\phi$ is in \textsc{cnf(2)} if it is a \textsc{cnf} and
every variable occurs at most twice (variables may occur positively
and negatively, and clauses may be arbitrary large). Johannsen showed
that the satisfiability problem and the nae-satisfiability problem for
\textsc{cnf(2)} formulas are complete for
$\Class{L}$~\cite{Johannsen04}. We extend this result and observe that
the logspace algorithms can be modified such that they solve
the corresponding maximization problem: Given a \textsc{cnf(2)}
formula $\phi$, they output the maximum
number of simultaneously satisfiable clauses. Combined with
Observation~\ref{observation:almost-lowerbound} we obtain:
\begin{theorem}\label{theorem:almost-sat(2)}
  $\PLang[k]{almost-nae-sat(2)}$ and $\PLang[k]{almost-sat(2)}$ are complete for $\Para\Class{L}$.
\end{theorem}

\begin{lemma (proof possibly in appendix)}\label{lemma:almost-sat(2)}
  There is a function in $\Class{FL}$ that maps {\normalfont\textsc{cnf(2)}}
  formulas $\phi$ to the maximum number of simultaneously satisfiable
  clauses of $\phi$.
\end{lemma (proof possibly in appendix)}
\begin{proof (possibly in appendix)}[Proof of Lemma~\ref{lemma:almost-sat(2)}]{
  We follow the proof by Johannsen~\cite{Johannsen04} and first count
  and remove
  all empty clauses (these can never be satisfied), then we represent
  $\phi$ as \emph{tagged graph} $G(\phi)$. Such a graph is a triple
  $(V,E,T)$ in which $V=\clauses(\phi)$ is the set of vertices,
  \[
    E=\bigl\{\{C_i,C_j\}\mid \exists x\in\vars(\phi) [x\in  C_i\wedge\neg x\in C_j]\bigr\}
  \] is a \emph{multiset} of undirected edges that connects
  clauses that contain complementary literals, and $T\subseteq V$ is a
  set $T=\{\,C_i\mid C_i\text { contains a pure literal}\,\}$ of
  tagged vertices (a literal is \emph{pure} if the negated literal is
  not present in the formula or, equivalently, if the occurrences of
  the literal's   variable are either all positive or all
  negated). Note that the 
  graph is a multigraph, i.\,e., if 
  clauses share multiple complementary literals, they
  are connected by multiple edges.

  Johannsen observed that the satisfiability problem of $\phi$ is
  equivalent to the following orientation problem of $G(\phi)$
  (Proposition~1 in~\cite{Johannsen04}): \emph{Can the edges of
    $G(\phi)$ be directed in such a way that there is no untagged
    sink?} The intuition is that tagged clauses can greedily be
  satisfied by setting the pure literal they contain, and that a
  variable $x\in\vars(\phi)$ can be used to satisfy exactly one of the
  two clauses it connects~--~orienting an edge $\{C_i,C_j\}$ as
  $C_i\rightarrow C_j$ thus means to set $x$ such that it satisfies
  $C_i$ but has no effect on $C_j$.

  Any connected component of $G(\phi)$ that contains a tagged vertex
  $v$ can be oriented in this way (just perform a depth-first search
  from $v$ and orient all edges towards the root of the dfs-tree). If
  a connected component contains a cycle, we can satisfy all vertices
  on that cycle by orienting it as directed cycle. Then we can
  virtually contract the cycle, tag the resulting vertex, and use the
  previous argument. Hence, Johannsen concluded~\cite{Johannsen04}:
  $\phi$ is satisfiable iff $G(\phi)$ does not contain a connected
  component without a tagged vertex that is a tree.  Since computing
  connected components and testing whether a component is a tree can
  be done in logarithmic space, it follows that $\Lang{sat(2)}\in\Class{L}$.

  If $\phi$ is satisfiable, the $\Class{FL}$ function that we
  wish to construct simply outputs $m$, the number of clauses. So 
  assume that $\phi$ is unsatisfiable. By the above argument,
  $G(\phi)$ then contains connected components $T_1,\dots, T_k$
  ($k\geq 1$) that are trees and that do not contain tagged vertices
  (these are the unsatisfiable cores of $\phi$). To make $\phi$
  satisfiable, we have to delete at least one clause per core, thus,
  we can satisfy at most $m-k$ clauses.

  On the other hand, deleting any clause $C$ in a tree $T_i$ will make
  all literals contained in~$C$ pure and, thus, will tag all neighbors
  of $C$ in $G(\phi)$. Hence, by deleting an arbitrary clause from
  $T_i$ we can make the remaining clauses of $T_i$ satisfiable. In
  conclusion, we can always satisfy at least $m-k$ clauses and, thus,
  we can output $m-k$ (taking into account the clauses removed in the
  preprocessing step).
}\end{proof (possibly in appendix)}

\begin{lemma (proof possibly in appendix)}\label{lemma:almost-nae-sat(2)}
  There is a function in $\Class{FL}$ that maps {\normalfont\textsc{cnf(2)}}
  formulas $\phi$ to the maximum number of simultaneously nae-satisfiable
  clauses of $\phi$.
\end{lemma (proof possibly in appendix)}
\begin{proof (possibly in appendix)}[Proof of Lemma~\ref{lemma:almost-nae-sat(2)}]{
  The proof is similar to the proof of
  Lemma~\ref{lemma:almost-sat(2)}: On input $\phi$, we first count and
  remove all empty and unit clauses (these can never be nae-satisfied). Then we construct a
  tagged graph $H(\phi)=(V,E,T)$ as follows:

  \begin{itemize}
    \item $V=\clauses(\phi)\cup\{\,d_x\mid\text{$x\in\vars(\phi)$
        appears positive and negative}\,\}$;
    \item \(
      E = \{\, \{C_i,C_j\} \mid \text{$C_i$ and $C_j$ contain the same literal}\,\}
      \cup
      \{\, \{C,d_x\} \mid \text{$x\in C$ or $\neg x\in C$} \,\}
    \);
    \item $T=\{\,C\mid \text{$C\in\clauses(\phi)$ contains a variable
        that does not occur in another clause}\,\}$.      
   \end{itemize}

   Johannsen observed that $\phi$ is nae-satisfiable iff \emph{the
     edges of $H(\phi)$ can be colored with two colors such that each
     untagged vertex is adjacent to edges of both
     colors}~\cite{Johannsen04}. (The intuition is that edges
   correspond to literals and colors represent truth values of these
   literals; tagged vertices can always be nae-satisfied with their
   private literal.)

   Since a tagged graph can be colored in the described way iff (i)
   every untagged vertex has degree at least two and (ii) every
   connected component without tagged vertices is \emph{not} a simple
   odd length cycle (Lemma~9 in~\cite{Johannsen04}), we get
   $\Lang{nae-sat(2)}\in\Class{L}$ (both criteria can easily be
   checked in logarithmic space).

   As in Lemma~\ref{lemma:almost-sat(2)}, if $\phi$ is satisfiable,
   the $\Class{FL}$ function we construct simply outputs $m$. So assume
   otherwise. Then there are connected components $O_1,\dots, O_k$ in
   $H(\phi)$ that do not contain a tagged vertex and that are simple
   odd length cycles (these are the nae-unsatisfiable cores of
   $\phi$). Clearly, any assignment can satisfy at most $m-k$ clauses.

   However, deleting an arbitrary clause $C$ from an odd cycle $O_i$
   will tag all the neighbors of $C$ (either the neighbor is another
   clause that now has a private variable, or it is a dummy vertex
   that now corresponds to a variable that occurs only once). Hence,
   we can satisfy at least $m-k$ clauses and can, thus, output $m-k$
   (taking into account the amount of clauses we have removed in
   the preprocessing step).  
}\end{proof (possibly in appendix)}

\subsection{Dual Parameterization for Formulas in Disjunctive Normal Form}
\label{section:dual-dnf}

Testing whether a \textsc{dnf} is satisfiable can be done in
polynomial time (even in $\Class{AC}^0$), in contrast, deciding
whether we can satisfy $k$ terms simultaneously (i.\,e.,
\Lang{max-dnf}) is $\Class{NP}$-complete~\cite{EscoffierP05}. In this
section we study \Lang{max-dnf} with a dual parameterization:
$\PLang[k]{almost-dnf}$ asks whether a given \textsc{dnf} $\phi$ has
an assignment that satisfies at least $m-k$ terms.

\begin{theorem}\label{theorem:almost-dnf}
  $\PLang[k]{almost-dnf}\in\Para\Class{AC}^0$.
\end{theorem}

The proof of the theorem boils down to the following reduction and the
subsequent lemma. Construct a \textsc{cnf} $\psi$ from $\phi$ by
simply negating every term, i.\,e., if
$(\ell_1\wedge\dots\wedge\ell_d)$ is a term in $\phi$, we add
$(\neg\ell_1\vee\dots\vee\neg\ell_d)$ as clause to $\psi$. Observe
that every assignment that satisfies a term in $\phi$ does \emph{not}
satisfy the correspond clause in $\psi$. Hence, there is an assignment
satisfying at least $m-k$ terms in $\phi$ if there is an assignment
that satisfies \emph{at most} $k$ clauses in $\psi$. In other words,
we have reduced $\PLang[k]{almost-dnf}$ to $\PLang[k]{min-sat}$.

\begin{lemma (proof possibly in appendix)}\label{lemma:min-sat}
  $\PLang[k]{min-sat}\in\Para\Class{AC}^0$.
\end{lemma (proof possibly in appendix)}
\begin{proof (possibly in appendix)}[Proof of Lemma~\ref{lemma:min-sat}]
  The following reduction from $\PLang[k]{min-sat}$ to
  $\PLang[k]{vertex-cover}$ by Marathe and Ravi~\cite{MaratheR96} is computable
  in $\Para\Class{AC}^0$ and is parameter-preserving. It takes an input
  $(\phi,k)$ and constructs a vertex cover instance $(G(\phi),k)$ as follows:
  The vertex set of $G(\phi)$ is $\clauses(\phi)$ and two clauses are
  connected by an edge if they contain complementary literals.
  Observe that any variable that occurs both, positively and
  negatively, satisfies at least one clause. In other words, every
  edge in $G(\phi)$ connects two clauses such that any assignment
  satisfies at least one of them. Therefore, we have
  $(\phi,k)\in\PLang[k]{min-sat}$ iff the edges of $G(\phi)$ can be
  covered by at most $k$ vertices, i.\,e., if
  $(G(\phi),k)\in\PLang[k]{vertex-cover}$.
  The claim follows as $\PLang[k]{vertex-cover}\in\Para\Class{AC}^0$,
  see Theorem 4.5 in~\cite{BannachST15}.
\end{proof (possibly in appendix)}

\section{Structural Parameterizations for Partial MaxSAT Variants}
\label{section:structure}

The most general incarnation of \Lang{max-sat} is the \emph{partially
  weighted version:} We are given a \textsc{cnf} $\phi$ and a weight
function
$\omega\colon\clauses(\phi)\rightarrow\mathbb{N}\cup\{\infty\}$, in
which we call clauses $C$ \emph{soft} if $\omega(C)<\infty$ and
\emph{hard} otherwise. The goal is to find among all assignments
$\beta\colon\vars(\phi)\rightarrow\{0,1\}$ that satisfy \emph{all} hard
clauses the one that \emph{maximises} the sum of the satisfied
soft clauses. We refer to the decision version, in which a
target sum is given, as \Lang{partial-max-sat}. Clearly, as in Observation~\ref{observation:almost-lowerbound}, the problem
restricted to a family of formulas $\Phi$ is at least as hard as the
satisfiability problem for $\Phi$, but we can even model the
independent set problem (find $k$ pairwise non-connected vertices in a
graph $G=(V,E)$) as a \textsc{2cnf} $\phi_{\text{is}}$ in which every
clause contains at most one positive literal and, thus, even
$\PLang[k]{partial-(2sat{$\cap$}horn)}$ is $\Class{W}[1]$-hard: 
\[
  \phi_{\text{is}} = 
  \underbrace{\textstyle\bigwedge_{\{u, v\}\in E}(\neg x_u\vee\neg x_v)}_{\text{hard clauses}}
  \wedge
  \underbrace{\textstyle\bigwedge_{v\in V}(x_v)}_{\text{soft clauses}}.
\]

The usual approach to identify tractable fragments of
\Lang{partial-max-sat} is to use \emph{structural parameters}, see~\cite{DellKLMM17} for an overview. Structural parameters are
defined over the \emph{incidence graph} of the input formula $\phi$,
which is the bipartite graph on vertex set
$\vars(\phi)\cup\clauses(\phi)$ that contains an edge between
$x\in\vars(\phi)$ and $C\in\clauses(\phi)$ if either $x\in C$ or
$\neg x\in C$.

Natural parameters are the \emph{vertex cover number}, the \emph{treedepth}, the
\emph{feedback vertex set number}, or the \emph{treewidth} of the
incidence graph. See Figure~\ref{figure:structural-parameters}
\iftcsautomove in the appendix\fi\ for an
overview of how these parameters are related. It is well-known that \Lang{partial-max-sat} is in
$\Para\Class{P}$ parameterized by any of these, which follows quite directly from
optimization versions of Courcelle's Theorem~\cite{Courcelle90} (see for instance
Chapter~17 in~\cite{HandbookSAT}).
In fact, by the parallel version of this
theorem~\cite{BannachT16} its follows that \Lang{partial-max-sat} lies
in $\Para\Class{AC}^{2\uparrow}$ if parameterized by \emph{both,} the
structural parameter and the solution size. However, this results in a
nonconstructive algorithm (i.\,e., we do not obtain an assignment)
and, as mentioned, works only if the solution size is a parameter,
too.

\begin{figure (possibly in appendix)}[htbp]
  \begin{tikzpicture}[
    parameter/.style = {
      font = \small,
      baseline
    },
    thispaper/.style = {
      font = \small\it
    }
    ]    
    \coloredclass{0}{mygreen}{80}{1}
    \classlabel{0}{$\Para\Class{TC}^0$}{$O(1)$}
    \node[parameter] (tc) at (0.2*\linewidth, 0.5) {twin cover};
    \node[parameter,thispaper] (vc) at (\linewidth/2, 0.5) {vertex cover};

    \coloredclass{1}{mygreen}{70}{2}
    \classlabel{2}{$\Para\Class{TC}^{0\uparrow}$}{$f(k)$}
    \node[parameter] (vi) at (0.6*\linewidth, 1.5) {vertex integrity};
    \node[parameter,thispaper] (td) at (0.6*\linewidth, 2.5) {treedepth};

    \coloredclass{3}{mygreen}{60}{1}
    \classlabel{3}{$\Para\Class{TC}^{1\uparrow}$}{$f(k)\log n$}    
    \node[parameter,thispaper] (fvs) at (0.75*\linewidth, 3+0.5) {feedback vertex set};

    \coloredclass{4}{mygreen}{50}{1}
    \classlabel{4}{$\Para\Class{TC}^{2\uparrow}$}{$f(k)\log^2 n$}    
    \node[parameter,thispaper] (tw) at (0.6*\linewidth, 4+0.5) {treewidth};

    \coloredclass{5}{myred}{100}{3}    
    \classlabel{7}{$\Class{W}[1]$}{intractable}

    \coloredclass{8}{myred}{80}{1}    
    \classlabel{8}{$\Para\Class{NP}$}{intractable}

    \node[parameter] (nd) at (0.34*\linewidth, 5+0.5) {neighborhood diversity};
    \node[parameter] (mw) at (0.2*\linewidth, 5+1.5)  {modularwidth};
    \node[parameter] (sd) at (\linewidth/2, 5+1.5)    {shrubdepth};
    \node[parameter] (oc) at (0.75*\linewidth, 7+1.5) {odd cycle transversal};
    \node[parameter] (cw) at (\linewidth/2, 5+2.5)    {cliquewidth};
    
    \graph[use existing nodes, edges = {semithick, ->, >={[round,sep]Stealth}}] {
      vc --[bend left=25] nd;
      vc -- tc --[bend right=20] sd;
      tc -- mw;
      vc -- vi -- td -- tw --[bend right] cw;
      td --[bend left=10] sd;
      vc --[bend right=40] fvs -- {tw, oc};
      nd -- {mw, sd} -- cw;
    };
    
    \node[             baseline, color=ba.blue, font=\sf\large] at (\linewidth/2, 10) {Structural Graph Parameter};
    \node[anchor=west, baseline, color=ba.blue, font=\sf\large] at (0,10)             {Complexity Class};
    \node[anchor=east, baseline, color=ba.blue, font=\sf\large] at (\linewidth,9.75)  {\parbox{4cm}{\flushright Parallel Time Using\\ $f(k)\cdot n^{O(1)}$ Processors}};
  \end{tikzpicture}  
  \caption{
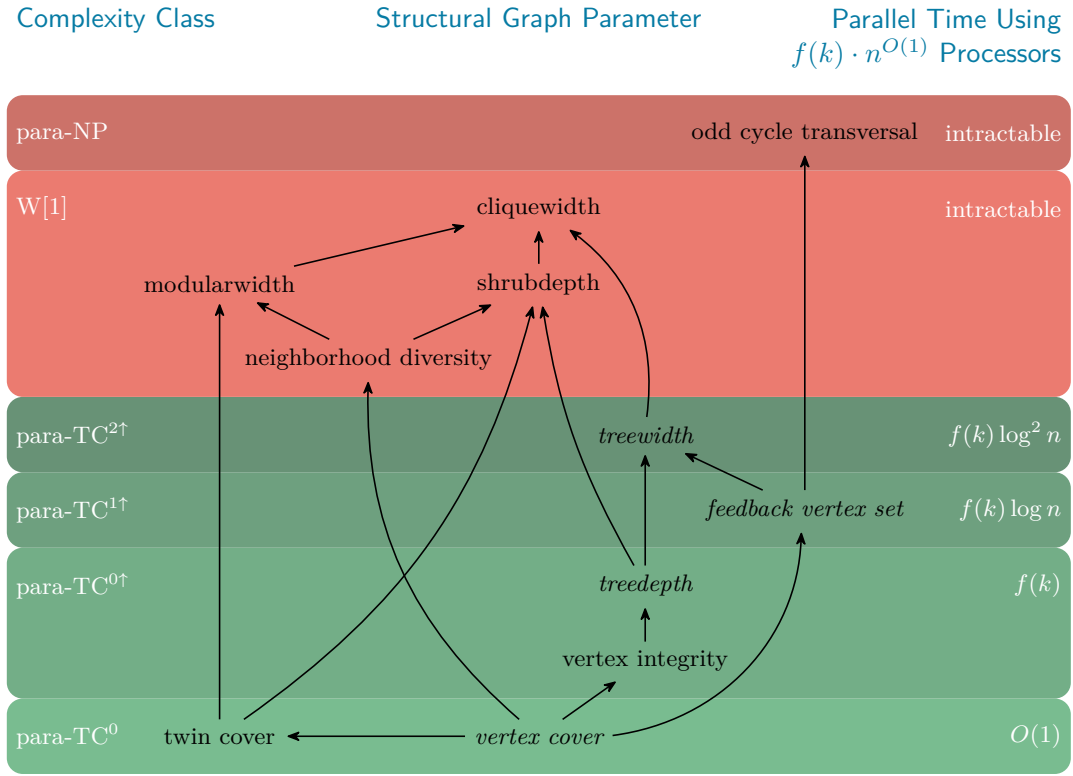
    A Hasse diagram of the major structural graph parameters. An arrow
    from $A$ to $B$ here means that for any graph $G$ the parameter
    $B$ is upper-bounded by a function in $A$. Each node corresponds
    to the complexity of $\Lang{partial-max-sat}$ parameterized by the
    corresponding value of the input's incidence graph. The emphasized
    entries are the results proven in
    Theorem~\ref{theorem:structural}. The \emph{neighborhood
      diversity} of a graph is the minimum number of partitions of a
    graph such that all vertices in the same partition are pairwise
    twins~--~the corresponding hardness result is proven in Lemma~3.5
    in~\cite{DellKLMM17}. The \emph{twin cover number} is a natural
    generalization of the vertex cover number in which only edges that
    do not connect twins have to be covered~\cite{Ganian11}. However, for bipartite
    graphs (as the incidence graph is) this is basically the same as
    the vertex cover number (the sole exception are isolated
    edges). Likewise, a parameterization by the size of an odd cycle
    transversal (a set whose deletion makes the graph bipartite) has
    no effect on bipartite graphs. The \emph{vertex integrity} is a
    parameter that closes the rather large gap between the vertex
    cover number and the treedepth~\cite{DrangeDH16}: It is the smallest value $k$ such
    that there is a vertex separator of size $k'\leq k$ and such that
    after the removal of that separator every component has size hat
    most $k-k'$. However, just to compute connected components of
    size~$O(k)$, circuits of constant depth require size roughly
    $O(n^k)$~\cite{BeameIP98} and, thus, this parameter is not suitable to break the
    barrier to $\Para\Class{TC}^{0}$. We refer the interested reader
    to~\cite{DellKLMM17} and the references therein for the definitions of the
    remaining $\Class{W}[1]$-hard parameters.
  }
  \label{figure:structural-parameters}
\end{figure (possibly in appendix)}

\clearpage

In the remainder of this section we develop handcrafted algorithms for
all four structural parameters that (i) work independently of the
solution size (it does \emph{not} have do be a parameter), (ii) work
with arbitrary weights, and (iii) are constructive in the sense that an
optimal assignment is output. The proof of the main result of this section,
Theorem~\ref{theorem:structural}, is presented in form of four
lemmas. Figure~\ref{figure:structural-parameters} \iftcsautomove
in the appendix\fi\ reveals intriguing connections
between these parameters (which can be partially ordered) to the degree of
parallelism we can achieve~--~a detail that is usually concealed in
the study of sequential $\Para\Class{P}$
algorithms.

\begin{theorem}\label{theorem:structural}
  $\PLang[\mathrm{vc}]{partial-max-sat}\in\Para\Class{TC}^0$,
  $\PLang[\mathrm{td}]{partial-max-sat}\in\Para\Class{TC}^{0\uparrow}$,
  $\PLang[\mathrm{fvs}]{partial-max-sat}\in\Para\Class{TC}^{1\uparrow}$,
  $\PLang[\mathrm{tw}]{partial-max-sat}\in\Para\Class{AC}^{2\uparrow}$.
\end{theorem}

\begin{lemma (proof possibly in appendix)}\label{lemma:structural-vc}
  There is a uniform family of
  constant-depth $\Class{TC}$ circuits of size $f(k)\cdot n^{O(1)}$
  that, on input $(\phi,\omega,k)$, either reports that the incidence
  graph of $\phi$ has no vertex cover of size~$k$, or that outputs the
  assignment of an optimal solution for \Lang{partial-max-sat} on $(\phi,\omega)$.
\end{lemma (proof possibly in appendix)}
\begin{proof (possibly in appendix)}[Proof of Lemma~\ref{lemma:structural-vc}]{

  First construct the incidence graph (which is easy in $\Class{AC}^0$)
  and then run the parallel version of the Buss kernel on it (with
  parameter~$k$)~\cite{BannachST15}.  If the kernelization algorithm
  produces a trivial no-instance (if it ``rejects''), then
  there is no vertex cover of size~$k$ and we may reject. Otherwise, we
  have a kernel with $k^2+k$ vertices and a set of at most~$k$
  vertices that were classified by the Buss kernel as being necessary
  for any vertex cover (i.\,e., high-degree vertices). Together we
  obtain a vertex cover $X$ of size at most $k^2+2k$.

  Let $V_1=\vars(\phi)\cap X$ and $V_2=\vars(\phi)\setminus V_1$; $S_1=\{\, C\in \clauses(\phi)\mid\omega(C)<\infty\,\}\cap X$ and
  $S_2=\{\, C\in \clauses(\phi)\mid\omega(C)<\infty\,\}\setminus S_1$,
  and define $H_1$ and $H_2$ analogously for the hard clauses. The
  circuit now runs the following three steps in sequence:
  \begin{enumerate}
  \item Brute-force (i.\,e., test in parallel) all possible partial assignments
    for the variables in $V_1$. Discard assignments that leave a
    clause in $H_2$ unsatisfied. 
  \item Guess (i.\,e., test in parallel) which clauses of $\tilde
    S_1\subseteq S_1$ shall be satisfied.
  \item Verify that the current partial solution can be extended to an
    assignment that satisfies $\tilde S_1\cup H_1$. Since $|\tilde
    S_1\cup H_1|\leq k^2+2k$, we can check if there is an assignment
    satisfying all clauses of this subformula using
    Theorem~\ref{theorem:max-circuit-sat}.
  \end{enumerate}

  Note that the first step already determines the truth value of all
  clauses in $S_2\cup H_2$. It remains to determine the sum of
  the weights of all satisfied clauses, which is easy in
  $\Class{TC}^0$. Observe that all three steps are constructive,
  i.\,e., at this point we have a list of roughly $2^{k^2+k}$
  assignments and their weights~--~we just have to output the one with
  the largest weight.   
}\end{proof (possibly in appendix)}

\begin{lemma (proof possibly in appendix)}\label{lemma:structure-td}
  There is a uniform family of $\Class{TC}$ circuits of depth $f(k)$
  and size $f(k)\cdot n^{O(1)}$ that, on input $(\phi,\omega,k)$,
  either reports that the treedepth of the incidence graph of $\phi$
  exceeds~$k$, or that outputs the assignment of an optimal solution
  for \Lang{partial-max-sat} on $(\phi,\omega)$.
\end{lemma (proof possibly in appendix)}
\begin{proof (possibly in appendix)}[Proof of Lemma~\ref{lemma:structure-td}]{
    A \emph{treedepth decomposition} of an undirected graph
    $G=(V,E_G)$ is a rooted forest $F=(V,E_F)$ (on the same vertex set)
    such that $G$ is a subgraph of the closure of $F$. The \emph{treedepth} of $G$ is the minimum depth any treedepth decomposition
    of $G$ must have, see~\cite{Sparsity} for a detailed introduction to these notations.

    A uniform family of $\Para\Class{FAC}^{0\uparrow}$ circuits is known that
    maps a pair $(G,k)$ either to $\bot$ (in which case the treedepth
    of $G$ exceeds~$k$) or to a treedepth decomposition $F$ of depth
    at most~$O(2^k)$, see Theorem~5 in~\cite{BannachT16}. Furthermore,
    if access to a depth-$f(k)$ treedepth decomposition is provided,
    $\Para\Class{FAC}^{0\uparrow}$ circuits can perform depth-first
    and breadth-first searches on $G$ and, thus, can compute
    connected components (Lemma~6 in~\cite{BannachT16}).

    Since $\Para\Class{FAC}^{0\uparrow}\subseteq\Para\Class{FTC}^{0\uparrow}$,
    we can assume that we have access to a depth-$k'$ treedepth
    decomposition $F$ of the incidence graph of $\phi$ (with $k'\in
    O(2^k)$), and that we can compute connected components in the
    incidence graph. Let us denote for a \textsc{cnf} $\psi$, a
    variable $x\in\vars(\psi)$, and $i\in\{0,1\}$, by $\psi|_{x\mapsto
      i}$ the formula obtained by deleting all clauses from $\psi$
    that are satisfied by setting $x$ to $i$, and by removing all
    remaining occurrences of~$x$ from the remaining clauses.  The
    claim is proven by running \textsf{DPLL} with a 
    variable selection heuristic based on~$F$. Clearly,
    the algorithm from Listing~\ref{algo:dpll} correctly solves
    \Lang{partial-max-sat}. 

    \input{dpll-td}

    We are left with the task of arguing that a circuit family of depth
    $f(k)$ and size $f(k)\cdot n^{O(1)}$ can implement this
    algorithm. All operations can be computed by
    $\Para\Class{TC}^{0\uparrow}$ circuits: Computing the connected
    components in line~\ref{line:cc} can be done in depth $f(k)$ since the
    treedepth is bounded; the sum of multiple binary numbers in
    line~\ref{line:sum} can be
    computed in constant depth (using threshold gates) by a result of
    Chandra, Stockmeyer, and Vishkin~\cite{ChandraSV84}; and all
    remaining operations are either simple arithmetic or the computation
    of projections.

    Since the depth of $F$ is
    $k'$, the recursion depth of the algorithm is $O(k')\subseteq
    O\big(2^{k}\big)$. Finally, since $F$ has at most
    $n+m=\left|\vars(\phi)\right|+\left|\clauses(\phi)\right|$ leaves, the total number of
    explored subformulas is bounded by $O\big(2^{2^k}(n+m)\big)$.

    We get the claim by adapting the algorithm such that it does not
    only return the maximum solution, but also the corresponding
    assignment. In detail, we assume a total order on the variables of
    $\phi$, i.\,e., $\vars(\phi)=\{x_1,\dots, x_n\}$, and represent an
    assignment $\beta\colon\vars(\phi)\rightarrow\{0,1\}$ as bit mask
    $\beta\in\{0,1\}^n$.  At the end of the recursion, i.\,e., in 
    lines~\ref{line:recusion-end-1} and~\ref{line:recusion-end-2}, we
    return an assignment $x\mapsto 0$ in the form of
    $\beta=0^n$. Getting such an assignment from the recursive call in
    line~\ref{line:recusion}, we can obtain a corresponding
    assignment $\beta_i$ by setting the bit corresponding to~$x$ to~$i$. Finally, after the recursion into connected components
    following line~\ref{line:cc}, we return the
    \emph{bitwise or}
    of all obtained assignments in line~\ref{line:sum} (note that,
    since the formula was disconnected, these assignments modified
    pairwise different bits).  }\end{proof (possibly in appendix)}

\begin{lemma (proof possibly in appendix)}\label{lemma:structure-fvs}
  There is a uniform family of $\Class{TC}$ circuits of depth
  $f(k)\cdot \log n$ and size $f(k)\cdot n^{O(1)}$ that, given $(\phi,\omega,k)$,
  either reports that $\phi$'s incidence graph has no
  size-$k$ feedback vertex set, or outputs the assignment of an optimal solution
  for \Lang{partial-max-sat} on $(\phi,\omega)$.
\end{lemma (proof possibly in appendix)}
\begin{proof (possibly in appendix)}[Proof of Lemma~\ref{lemma:structure-fvs}]{

    A \emph{tree decomposition} of an undirected graph $G=(V_G,E_g)$ is a tuple $(T,\iota)$, where $T=(V_T, E_T)$ is a tree and
    $\iota\colon V_T\rightarrow 2^{V_G}$ a mapping from nodes of $T$
    to subsets of vertices of $G$, which we call \emph{bags}. A tree
    decomposition has to satisfy the following constraints, see
    Chapter~7 in~\cite{CyganFKLMPPS15} for a detailed introduction:

    \begin{itemize}
    \item The set $\{\,x\mid v\in\iota(x)\,\}$ is non-empty and
      connected in $T$ for every $v\in V_G$.
    \item For every $\{u,v\}\in E_G$ there is a $y\in V_T$ with
      $\{u,v\}\subseteq\iota(y)$.
    \end{itemize}
    
    The \emph{width} of $(T,\iota)$ is the size of the largest bag minus one, and
    the \emph{treewidth} of $G$ is the minimum width any tree
    decomposition of $G$ must have.

    Clearly, a graph with a feedback
    vertex set $X$ of size at most $k$ has treewidth at most $k+1$:
    Remove $X$ from $G$ and obtain a tree, then consider the tree as
    tree decomposition and add $X$ to every bag. Hence, by
    Lemma~\ref{theorem:fvs} (and since
    $\Para\Class{FL}^{\uparrow}\subseteq\Para\Class{TC}^{1\uparrow}$),
    we can either conclude that the incidence graph of $\phi$ has no
    feedback vertex set of size~$k$, output this and stop; or we can
    compute a tree 
    decomposition of width at most $k+1$.  In order to make the
    description of the following dynamic program simpler, we bring
    $(T,\iota)$ into a standard form called \emph{balanced nice tree
      decomposition.} In this form, $T$ is a rooted tree of depth
    $f(k)\cdot O(\log n)$ such that every node $x$ has one of the
    following types:

    \begin{description}
    \item[Leaf Nodes] They have no children.
    \item[Introduce Nodes] They have exactly one child~$y$ with
      $\iota(x)=\iota(y)\cup\{v\}$ for a $v\in V_G$.
    \item[Forget Nodes] They have exactly one
      child $y$ and there is vertex $v\in V_G$ with
      $\iota(x)=\iota(y)\setminus\{v\}$.
    \item[Join Nodes] They have exactly two
      children $y$ and $z$ with
      $\iota(x)=\iota(y)=\iota(z)$.
    \end{description}

    A uniform family of $\Para\Class{FNC}^{1\uparrow}$ circuits that
    maps arbitrary width-$k$ tree decomposition to balanced nice tree
    decompositions of width at most $8k+3$ is known (Lemma~9
    in~\cite{BannachT16}). Since
    $\Para\Class{FNC}^{1\uparrow}\subseteq\Para\Class{TC}^{1\uparrow}$
    we may assume that $(T,\iota)$ is in this special form. Let us, to
    keep the notation intuitive, denote the size of the largest bag of
    the transformed decomposition with $k$ (even though it did, of
    course, grow a little).

    Our task is to describe a family of
    $\Para\Class{FTC}^{1\uparrow}$ circuits that obtains as input a
    tuple $\big((\phi,\omega,T,\iota),k\big)$ (where $(T,\iota)$ is a
    width-$(k+1)$ tree decomposition of the incidence graph of~$\phi$ and
    $k$ is the parameter) and outputs an assignment that satisfies all
    hard clauses while maximising the sum of the weights of satisfied
    soft clauses (or detects that such an assignment does not
    exist). The idea of the following algorithm is a dynamic program
    that bubbles up the tree decomposition, assigning
    \emph{configuration sets} to the nodes of~$T$. We think of $T$ as
    being layered (with $f(k)\cdot\log n$ layers). Thus, all we have
    to do is to design $\Class{TC}$ circuits of depth $f'(k)$
    (independent of $n$) and size $f'(k)\cdot n^{O(1)}$ for some
    computable function $f'$, which compute the configuration sets of
    a node in $T$, given the configuration sets of its children.
    
    Assume that there is a total order on the vertices of the incidence
    graph (for instance, take the lexicographical order induced by
    $\phi$). A \emph{configuration} is a triple $(\mu,\sigma,\beta)$,
    where $\mu\in\{0,1\}^k$ is a bit mask, $\sigma\in\mathbb{N}$ a
    weight, and $\beta\in\{0,1\}^n$ another bit mask (for a node $x\in V_T$ we interpret the
    bit masks as $\mu\colon\iota(x)\rightarrow\{0,1\}$ and
    $\beta\colon\vars(\phi)\rightarrow\{0,1\}$). For instance, say we
    have $\vars(\phi)=\{x_1,\dots,x_{100}\}$ and assume the incidence
    graph has width $6$. We encode an assignment $\beta$
    as bit mask $\beta\in\{0,1\}^{100}$ with the $i$th bit set iff $x_i$ is
    assigned to~$1$. For a node $x\in V_T$ of the tree decomposition we
    represent the local information
    $\mu\colon\iota(x)\rightarrow\{0,1\}$ as another bit mask
    $\mu\in\{0,1\}^5$, where the $i$th bit corresponds to the
    information stored for the lexicographical $i$th element of $\iota(x)$,
    e.\,g., if $\iota(x)=\{x_1,x_3,x_{42}\}$ we would store the
    information at the following positions:
    $\bigl(\begin{smallmatrix}
    \text{\scriptsize\it bit position:} & 1   & 2   & 3     & 4 & 5 \\
    \text{\scriptsize\it information:}  & x_1 & x_3 & x_{42} & 0 & 0
   \end{smallmatrix}\bigr).$ Note that we always set unused positions
   to the default value $0$. During the execution of the dynamic
   program, we will encounter new nodes of the tree decomposition that
   differ by at most one element of the previous one (i.\,e., we
   introduce a vertex to the bag). To reuse a previous $\mu$, we have
   to \emph{shift} its content accordingly. For instance, if $y$ is
   another bag with $\iota(y)=\iota(x)\cup\{x_{2}\}$, then for $y$ we
   would store the corresponding data at the following positions:
    $\bigl(\begin{smallmatrix}
    \text{\scriptsize\it bit position:} & 1 & 2 & 3 & 4 & 5 \\
    \text{\scriptsize\it information:}  & x_1 & x_2 & x_{3} & x_{42}  & 0
   \end{smallmatrix}\bigr)$, shifting data to the right from
   position~2 ongoing.

    We say two configurations
    $(\mu_1,\sigma_1,\beta_1)$ and $(\mu_2,\sigma_2,\beta_2)$ are
    \emph{equivalent} if $\mu_1=\mu_2$. Furthermore, a configuration
    is \emph{better} than another if $\mu_1=\mu_2$ and
    $\sigma_1>\sigma_2$. A \emph{configuration set} is a set of
    pairwise non-equivalent configurations. Note that such a set
    contains at most $2^k$ elements. For a node $x\in V_T$, a
    configuration $(\mu,\sigma,\beta)$ fulfills the following
    invariant:
    
    \begin{enumerate}
    \item For all $v\in\iota(x)\cap\vars(\phi)$ we have
      $\beta(v)=\mu(x)$.
    \item For all
      $C\in\iota(x)\cap\clauses(\phi)$ we have $\beta\models C$ iff
      $\mu(C)=1$.
    \item The assignment $\beta$ satisfies all
      hard clauses in the subtree rooted at $x$.
    \item The sum of the
      weights of soft clauses satisfied by $\beta$ in the subtree
      rooted at $x$ is $\sigma$.
    \end{enumerate}

    Since \nodetype{Leaf Nodes} have no children, there is not much to
    do for them. The configuration set contains a single configuration 
    $(0^k,0,0^n)$.

    The circuit for \nodetype{Introduce Nodes}~$x$ obtains as input a configuration set $S$ and an
    introduced vertex~$v$, and outputs the following configuration set
    $S'$: If $v\in\clauses(\phi)$, construct for every
    $(\mu,\sigma,\beta)\in S$ a single new configuration
    $(\mu',\sigma',\beta)\in S'$ obtained by shifting~$\mu$
    according to the ordering of the vertices in $\iota(x)$, setting
    $\mu'(v)=1$ iff $\beta\models v$ (and $\mu'(v)=0$ otherwise), and
    setting $\sigma'=\sigma+\omega(v)$ if $v$ is a satisfied soft
    clause, otherwise setting $\sigma'=\sigma$.
    If $v\in\vars(\phi)$, we add for
    every $(\mu,\sigma,\beta)\in S$ configurations
    $(\mu_0,\sigma_0,\beta_0)$ and $(\mu_1,\sigma_1,\beta_1)$ to~$S'$,
    where we obtain $(\mu_i,\sigma_i,\beta_i)$ for $i\in\{0,1\}$ from
    $(\mu,\sigma,\beta)$ by shifting~$\mu$, by setting $\mu_i(v)=i$,
    and by updating $\beta_i$ appropriate. Subsequently, we check for
    every clause $C\in \iota(x)$ with $\mu(C)=0$ whether
    $\beta_i\models C$, in which case we set $\mu_i(C)=1$ and update
    $\sigma_i$ as needed.

    For \nodetype{Forget Nodes} $x$ that forget
    a vertex $v$, we construct a new configuration set $S'$ from~$S$
    by initially setting $S'=S$. Then in all configuration we remove
    $v$ from $\mu$ by setting the corresponding bit to $0$ and by
    shifting~$\mu$ according to the new ordering in $\iota(x)$. If $v$
    was a hard clause, we remove all configurations with $\mu(v)=0$
    (i.\,e., configurations in which $v$ was not satisfied). Finally,
    $S'$ may now contain some equivalent configurations, in which case
    we keep just the one that is best.

    When joining two bags in a \nodetype{Join Node} $x$, we obtain two configuration sets $S_1$
    and $S_2$ as input, and have to construct a new configuration set
    $S'$ that fulfils the invariant. Recall that the children of $x$
    have the same bag as $x$. Define for every
    $X\subseteq\vars(\phi)\cap\iota(x)$ and $i\in\{1,2\}$:    
    \[
      S_i^X = \bigl\{
      (\mu,\sigma,\beta)\in S_i
      \mid
      \text{for all }v\in\vars(\phi)\cap\iota(x)\text{ we have }\mu(v)=1\Leftrightarrow v\in X\bigr\}.
    \]
    Build the sets $S'(X)$ by joining $S_1^X$ and
    $S_2^X$ in the following sense: Take every configuration
    $(\mu_1,\sigma_1,\beta_1)\in S_1^X$ and every
    $(\mu_2,\sigma_2,\beta_2)\in S_2^X$ and build the new
    configuration:
    \[
      (\mu_1\curlyvee \mu_2,
      \sigma_1+\sigma_2-\epsilon,
      \beta_1\curlyvee\beta_2).
    \]    
    Here, ``$\curlyvee$'' is the
    \emph{bitwise or} operation and $\epsilon$ the sum of soft
    clauses in $\iota(x)$ that are satisfied by both assignments,
    i.\,e., by $\beta_1$
    \emph{and}~$\beta_2$. The configuration set~$S'$ is obtained by,
    firstly, collecting all $S'(X)$ and, secondly, by removing
    equivalent 
    configurations from it.

    A standard induction shows that, after the process has
    finished, any configuration stored in the root bag of $T$ contains
    an assignment $\beta$ that satisfies all hard clauses.
    Furthermore, the one with maximum $\sigma$ corresponds to an
    optimal solution for \Lang{partial-max-sat}.

    The statement follows as all four operations can be implemented by
    $\Class{TC}$ circuits of depth $f(k)$ and size
    $f(k)\cdot n^{O(1)}$. For the part that manipulates $\mu$, this
    follows as we have at most $2^k$ configurations and since
    $|\mu|=k$. Operations modifying $\sigma$ just have to perform
    simple addition and subtraction, which is possible even in a
    constant number of $\Class{TC}$ layers of polynomial
    size. Regarding~$\beta$, we perform only trivial bit projections.
  }\end{proof (possibly in appendix)}

\begin{lemma (proof possibly in appendix)}\label{lemma:structure-tw}
  There is a uniform family of $\Class{AC}$ circuits of depth
  $f(k)\cdot \log^2 n$ and size $f(k)\cdot n^{O(1)}$ that, given $(\phi,\omega,k)$,
  either reports that the treewidth of the incidence graph of $\phi$
  exceeds~$k$, or outputs the assignment of an optimal solution
  for \Lang{partial-max-sat} on $(\phi,\omega)$.
\end{lemma (proof possibly in appendix)}
\begin{proof (possibly in appendix)}[Proof of Lemma~\ref{lemma:structure-tw}]{
    An optimal tree decomposition can be computed in
    $\Para\Class{FAC}^{2\uparrow}$~\cite{BannachT16}. Afterwards the proof is
    equivalent to the proof of Lemma~\ref{lemma:structure-fvs}.
}\end{proof (possibly in appendix)}

\section{Conclusion and Outlook}
\label{section:conclusion}

We presented a comprehensive list of parallel fixed-parameter
algorithms for variations of \Lang{max-sat}. 
Table~\ref{table:overview} on page~\pageref{table:overview} offers an
overview of the results and the used techniques. As highlight we
presented the first parallel algorithms for
$\PLang[k]{almost-nae-2sat}$ and $\PLang[k]{almost-2sat}$, which
implies parallel fpt-algorithms for various problems such as the odd
cycle transversal problem.

The central method for
proving that the latter problem is fixed-parameter tractable -- the
iterative compression method -- seems to be inherently
sequential. Interestingly, our parallel algorithm builds on another
method that seems inherently sequential in general, namely the
computation of maximum flows. However, using properties of
the Hochbaum network allowed us to break the computation of a maximum
flow into a series of small flow computations, which we then can
perform in parallel using fpt-many parallel processing units.

We remark that from a complexity-theoretic point of view,
$\PLang[k]{almost-2sat}$ is a harder problem than
$\PLang[k]{almost-nae-2sat}$ as the former is easily
seen to be hard for $\Para\Class{NL}$ while
the latter is easily seen to lie in
$\Para\Class{WL}$ (see~\cite{ElberfeldST15} for a discussion of these
classes), which suggests that the problems have different
complexity. As open problem we thus leave the question of whether
$\PLang[k]{almost-nae-2sat} \in \Para\Class{L}^\uparrow$ holds (which
would imply that the odd cycle transversal problem lies in this class,
too). While we know of no complexity-theoretic assumption that would
contradict this, our proofs make heavy use of finding augmenting paths
in networks and these networks seem to be inherently directed.


\clearpage
\bibliography{main}

\clearpage
\appendix

\tcsautomovefurtherappendix{
 \section{Appendix: Figures and Tables}
 \label{appendix:tables}
} 

 \clearpage

\tcsautomoveproofappendix{
  \section{Technical Appendix: Missing Proofs}
  \label{appendix:proofs}
  For the reader's convenience, we restate the claims of the different
  lemmas and theorems whose proofs have been moved to this appendix.
}

\end{document}

%% file: dpll-td.tex
\begin{lstlisting}[
  style=pseudocode,
  backgroundcolor=,
  label=algo:dpll,
  float=htpb,
  caption={An algorithm that outputs an optimal
    $\Lang{partial-max-sat}$ solution on  input of a weighted
    \textsc{cnf} $(\phi,\omega)$ and a depth-$k'$ treedepth
    decomposition~$F$ of the incidence-graph of $\phi$.} 
]
function $\mathsf{DPLL\text-TD}(\phi,\omega,F)$
   if $\phi$ contains a hard empty clause then return $-\infty$ $\label{line:recusion-end-1}$
   if $\phi$ is empty or contains only empty soft clauses then return $0$ $\label{line:recusion-end-2}$
   if the incidence graph of $\phi$ is unconnected then
      $\phi_1,\dots,\phi_{\ell} \gets$ the connected subformulas $\label{line:cc}$
      for $i \in \{1,\dots,\ell\}$ pardo
         $\sigma_i \gets \mathsf{DPLL\text-TD}(\phi_i,\omega,F)$
      return $\sum_{i=1}^{\ell}\sigma_i$  $\label{line:sum}$
   else
      $x \gets $ the variable in $\vars(\phi)$ closest to the root of $F$
      for $i \in \{0,1\}$ pardo
         $\delta_i \gets$ the sum of soft clauses satisfied by setting $x$ to $i$
         $\sigma_i \gets \mathsf{DPLL\text-TD}(\phi|_{x\mapsto i},\omega,F)$ $\label{line:recusion}$
      return $\max(\sigma_0+\delta_0, \sigma_1+\delta_1)$
\end{lstlisting}